\documentclass[10pt]{iopart}
\usepackage{iopams}
\usepackage{graphicx}
\usepackage{setstack}
\eqnobysec

\newcommand{\half}{\case{1}{2}}
\newcommand{\bk}{\bi{k}}
\newcommand{\nhat}{\hat n}
\newcommand{\adkisig}{a^{\dagger}_{\bk i \sigma}}
\newcommand{\akisig}{a^{\phantom\dagger}_{\bk i \sigma}}
\newcommand{\aadkisig}{a^{\dagger}_{k i \sigma}}
\newcommand{\aakisig}{a^{\phantom\dagger}_{k i \sigma}}

\newcommand{\cdisig}{c^{\dagger}_{i\sigma}}
\newcommand{\cisig}{c^{\phantom\dagger}_{i\sigma}}
\newcommand{\adkm}{a^{\dagger}_{\bk m}}
\newcommand{\akm}{a^{\phantom\dagger}_{\bk m}}

\newcommand{\cm}{c^{\phantom\dagger}_{m}}
\newcommand{\fd}{f^{\dagger}}
\newcommand{\fpd}{f^{\phantom\dagger}}
\newcommand{\fo}{FO}
\newcommand{\co}{CO}
\newcommand{\stc}{SC}
\newcommand{\lmtwo}{LM$^{SU(2)}$}
\newcommand{\lmfour}{LM$^{SU(4)}$}
\newcommand{\tktwo}{T_K^{SU(2)}}
\newcommand{\tkfour}{T_K^{SU(4)}}
\newcommand{\mub}{\mu_B}
\newcommand{\cd}{c^{\dagger}}

\newcommand{\cnod}{c^{\phantom\dagger}}

\newcommand{\ad}{a^{\dagger}}

\newcommand{\anod}{a^{\phantom\dagger}}

\begin{document}
\title{Renormalization group study of capacitively coupled double quantum dots}
\author{Martin R Galpin\dag, David E Logan\dag\ and H R Krishnamurthy\ddag}
\address{\dag\ Oxford University, Physical and Theoretical Chemistry
Laboratory, South   
Parks Road, Oxford OX1 3QZ, UK.}
\address{\ddag\ Department of Physics, IISc, Bangalore 560 012, and
JNCASR, Jakkur, Bangalore, India.}
\begin{abstract}
The numerical renormalization group is employed to study a double
quantum (DQD) dot system consisting of two equivalent single-level 
dots, each coupled to its own lead and with a mutual capacitive coupling embodied in an interdot interaction $U'$, in addition to the intradot Coulomb interaction $U$.
We focus on the regime with two electrons on the DQD, and the evolution of the system
on increasing $U'/U$. The spin-Kondo effect arising for $U'=0$ ($SU(2) \times SU(2)$) is found to persist robustly with increasing $U'/U$, before a rapid but continuous crossover to (a) the $SU(4)$ point $U'=U$ where charge and spin degrees of freedom are entangled and the Kondo scale strongly enhanced; and then (b) a charge-Kondo state, in which a charge-pseudospin is quenched on coupling to the leads/conduction channels. A quantum phase transition of Kosterlitz-Thouless type then occurs from this Fermi liquid, strong coupling (SC) phase, to a broken symmetry, non-Fermi liquid charge ordered (CO) phase at a critical $U'_{c}$.
Our emphasis in this paper is on the structure, stability and flows between
the underlying RG fixed points, on the overall phase diagram in the ($U,U'$)-plane and evolution of the characteristic low-energy Kondo scale inherent to the SC phase; and on static physical properties such as spin- and charge-susceptibilities (staggered and uniform), including universality and scaling behaviour in the strongly correlated regime. Some exact results for associated Wilson ratios are also obtained.

\end{abstract}

\vspace{28pt plus 10pt minus 18pt} \noindent{\small\rm Published as: Martin R Galpin et al 2006 {\it J. Phys.: Condens. Matter} {\bf 18} 6545-6570\\ Journal of Physics: Condensed Matter \copyright\ 2006 IOP Publishing Ltd. \par}

\maketitle
\section{Introduction}
\label{sec:intro}

  The advent of semiconducting quantum dots has stimulated in recent years 
a remarkable renewal of interest in the Kondo effect~\cite{kouw}, a central 
paradigm of condensed matter science more traditionally associated with bulk 
systems~\cite{hewson}.
At its simplest the classic spin-Kondo effect arises~\cite{GG,Cronen,vdw},
a single spin in an odd-electron dot being quenched by coupling to the low-energy degrees of freedom of the conduction electrons in a metallic lead;
as manifest strikingly in e.g.\ the unitarity limit for the zero-bias
conductance at low temperatures~\cite{vdw}. 
There is naturally considerable interest in coupled quantum dots at present,
notably double dot (DQD) systems which are under active investigation both 
theoretically [6-18]
and experimentally [19-27].
In addition to their possible utility in mesoscopic devices, circuitry and information processing, such systems offer the prospect of novel strongly correlated electron states, reflecting the inherent relevance of both spin \emph{and} orbital degrees of freedom.

 Probably the simplest DQD system is that in which two equivalent semiconducting
dots, each in effect consisting of a single level and separately 
connected to their own leads/conduction channels, are mutually coupled by a 
capacitive interaction embodied in an \emph{inter}dot interaction
$U'$ in addition to the usual intradot interaction $U$. 
The DQD itself can contain up to $n=4$ electrons, controlled by varying 
the dot levels (e.g.\ by application of suitable gate voltages). The problem
has been studied quite extensively in the $n=1$-electron sector~\cite{poh,Boese,borda}, where the degenerate DQD states $(n_{L},n_{R}) = (1,0)$ and $(0,1)$ dominate
and lead to a beautiful example of the $SU(4)$ Kondo effect involving spin and 
orbital degrees of freedom~\cite{Boese,borda}.

  In the present paper by contrast we report an in-depth study of the $n=2$-electron regime, and its evolution as the ratio $U'/U$ of interdot to intradot interaction strengths is increased. We uncover a rich range of physical behaviour, qualitatively quite distinct from the $n=1$ sector. For $U' =0$ the dots are obviously uncoupled, 
the DQD states $(n_{L},n_{R}) = (1,1)$ dominate and the normal spin-Kondo effect 
($SU(2) \times SU(2)$) prevails. That essential behaviour is found to persist on increasing $U'/U$, until close to $U'/U =1$ where we find a rapid but continuous 
crossover to the $SU(4)$ Kondo state occurring at $U'/U =1$. Here the six DQD states $(1,1), (2,0)$ and $(0,2)$ are degenerate, the spin and charge degrees of freedom are entangled, and the characteristic low-energy Kondo scale is markedly enhanced. 
On increasing $U'/U$ just beyond the $SU(4)$ point the system then enters a charge-Kondo state, likewise a Fermi liquid state, where a charge pseudospin arising from the DQD states $(2,0)$ and $(0,2)$ is Kondo quenched on coupling to the conduction channels. 

  That Kondo quenching is however fragile: for only a small increase in $U'/U$ the charge-pseudospin tunneling is suppressed, the associated Kondo scale collapses rapidly but continuously, and the system undergoes a Kosterlitz-Thouless quantum phase transition at a critical $U'_{c}$ from the Fermi liquid phase to a 
charge ordered phase, a broken-symmetry non-Fermi liquid state with $\ln 2$ residual entropy reflecting the degenerate, unquenched $(2,0)$ and $(0,2)$ DQD configurations. Here we study the problem using Wilson's numerical renormalization group (NRG) approach~\cite{wilson,kww1,kww2} as the natural method of choice. A preliminary account of some of the results has been given in a recent letter~\cite{ddprl}.

  The paper is organised as follows. The model is introduced in 
section~\ref{sec:model}, and the essential steps of the NRG procedure for it
are outlined in section~\ref{sec:nrg}. RG fixed points, their stability and
flows between them in different parameter regimes, are discussed in 
section~\ref{sec:fixedpoints}, with particular attention given to the 
charge ordered (CO) fixed point that controls the CO phase. 
In section~\ref{sec:fpa} a number of analytical results are 
obtained by considering how the leading corrections to the two stable fixed
points, strong coupling (SC) and CO, affect flows in their vicinity. For the 
SC phase in particular we consider the static spin susceptibility, uniform and staggered charge susceptibilities, and the linear coefficient of specific heat; 
and obtain a number of exact results for associated Wilson ratios.
Numerical results are presented in section~\ref{sec:numrestherm}. We
begin by establishing the phase diagram in the ($U,U'$)-plane,
together with the evolution of the Kondo scale in the various regimes of the SC phase.
Detailed results for the $T$-dependence of static properties are then given,
notably the `impurity' entropy, spin- and charge-susceptibilities.
The important issues of scaling and universality in the SC phase are likewise considered. Results for (and issues relating to) single-particle dynamics and the linear differential conductance, will be given in a subsequent paper~\cite{next}.
\section{Model}
\label{sec:model}
\begin{figure}
\centering\includegraphics[height=3cm]{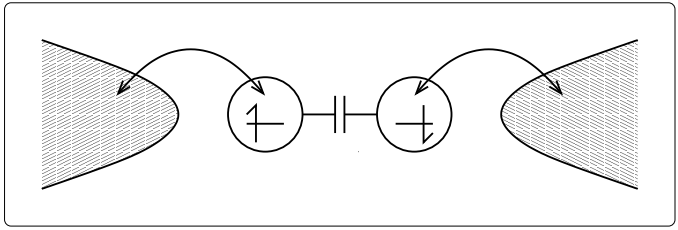}
\caption{\label{fig:fig1} Schematic representation of the double dot system}
\end{figure}

We model the capacitively coupled double dot system by a pair of correlated Anderson impurities, connected additionally  by an interimpurity Coulomb interaction $U'$ as shown schematically in \fref{fig:fig1}. In standard notation, the Hamiltonian can be written as $H=H_0+H_\mathrm{int}$ where
\numparts
\begin{eqnarray}
\label{eq:hnought}
H_0=\sum_{i,\bk,\sigma}\epsilon_\bk\adkisig\akisig+\sum_i\epsilon\nhat_i+\sum_{i,\bk,\sigma}V(\adkisig\cisig + \mathrm{h.c.}),\\
\label{eq:hint}
H_\mathrm{int}=\sum_i U\nhat_{i\uparrow}\nhat_{i\downarrow} + U'\nhat_L\nhat_R
\end{eqnarray}
\endnumparts
and $\hat n_i=\sum_\sigma\hat n_{i\sigma}=\sum_{\sigma}\cdisig\cisig$ is the number 
operator for dot $i\in\{L,R\}$ (referring to the `left' and `right' leads/channels).
The first two terms of $H_0$ describe the isolated lead and dot levels respectively, 
while the third term contains the one-electron hopping processes between each dot and its associated lead. $H_\mathrm{int}$ contains the interaction terms that render the problem non-trivial; here we include both an intradot Coulomb repulsion $U$ between the $\sigma=\uparrow$- and $\downarrow$-spin electrons on each dot, and the interdot $U'$ that embodies the capacitive coupling of the L and R channels. 

When $U'=0$, the Hamiltonian (\ref{eq:hnought},{\it b}) is simply a doubled version of the single-impurity Anderson model, the spin-Kondo physics of which is now well-understood \cite{hewson}. As we show in due course, the additional interdot correlations introduced by $U'>0$ tend to disfavour this spin-Kondo behaviour, leading to more complex states in which both spin and orbital degrees-of-freedom can be important. It is for example easy to show that when $U'=U$ the model maps onto the  $SU(4)$ Anderson model \cite{hewson,Boese}
\begin{equation}
\label{eq:hsufour}
\fl 
H=\sum_{\bk,m}\epsilon_\bk \adkm\akm + \sum_m\epsilon\hat n_m + \sum_{\bk,m} V(\adkm\cm + \mathrm{h.c.})+\sum_{m,m'}U\hat n_m\hat n_{m'}
\end{equation}
with spin and channel indices combined into a single quantity $m=(i,\sigma)$. The behaviour at this $SU(4)$ point is certainly quite different from the $SU(2)$ spin-Kondo physics of $U'=0$ (albeit that the associated stable fixed points are fundamentally of the same type in this case), and one would like to understand how the system evolves between the two with increasing $U'$. The situation beyond $U'=U$ is also of considerable interest: here one naturally expects charge degrees-of-freedom to become more important than spin, as indeed will be seen vividly throughout this paper.

To obtain a rough idea of its parameter regimes, it is first convenient to look at the model in the `atomic' limit ($V=0$, where the dots are disconnected from the leads). Here the isolated dot Hamiltonian is
\numparts
\begin{eqnarray}
\label{eq:ddot1}
H_{D} =\sum_{i =L,R}(\epsilon\nhat_i + U\nhat_{i\uparrow}\nhat_{i\downarrow}) + U'\nhat_L\nhat_R \\
\label{eq:ddot2}
~~~~~\equiv \epsilon \hat{n} + \case{1}{2}U'\hat{n}(\hat{n}-1) + 
[U-U']\sum_{i =L,R}\nhat_{i\uparrow}\nhat_{i\downarrow}
\end{eqnarray}
\endnumparts
with $\hat{n} = \sum_{i}\hat{n}_i$ the total DQD number operator.
For fixed interactions $U$ and $U'$, consider progressively lowering the energies of the dot levels $\epsilon<0$ (via e.g.\ a suitable gate voltage).
With increasing $|\epsilon|$, the total ground-state occupancy of the dots takes the form of the familiar Coulomb-blockade `staircase', the occupancy increasing stepwise from \(n=1\) to \(n=4\). The points of discontinuity in the staircase correspond of course to degeneracy of the $n$- and $(n+1)$-occupied configurations, and are easily shown to lie at $|\epsilon|=\min(U,U')$, $U'+\max(U,U')$ and $U+2U'$. The $n=1$ sector of the full DQD Hamiltonian (\ref{eq:hnought},{\it b}) has been studied previously \cite{borda}. In this paper by contrast we focus on the $n=2$ regime,
$\min(U,U') < |\epsilon| < U' + \max(U,U')$, and the rich range of physical behaviour arising therein. In particular we consider the midpoint, $|\epsilon|=\case{1}{2}U+U'$, at which the model is particle-hole symmetric; stressing at the outset that 
the $ph$-symmetric point is generically representative of the $n=2$ regime
(movement away from the symmetric point simply generates potential scattering in the two leads, of both equal sign and magnitude, and does not alter the essential physics of the problem). 

With $|\epsilon| = \case{1}{2}U+U'$ henceforth, there remains of course the whole of the ($U,U'$) plane to explore. In the atomic limit, this is clearly divided by the line $U'=U$. For any non-zero $U$, the two electrons occupy different dots in the ground-state when $U'<U$, but share the same dot when $U'>U$; i.e.\ labelling the states by their individual dot occupation numbers $n_i$, the ground state is $(n_L,n_R)=(1,1)$ for $U'<U$, but $(2,0)$ or $(0,2)$ for $U'>U$, as follows directly from
(\ref{eq:ddot2}). A quantum phase transition, albeit of a rather trivial kind, thus occurs in the atomic limit. But does a non-trivial quantum phase transition  arise
when the dots are connected to their leads? In the remainder of the paper we answer this, and associated questions, by analysing the model within the powerful framework of the NRG \cite{wilson,kww1,kww2}.
\section{The Numerical Renormalization Group}
\label{sec:nrg}
The NRG procedure for the DQD system is naturally quite similar to that of the single-impurity Anderson model. We thus refer the reader to \cite{kww1} for further details, but outline the essential steps here in order to point out the differences that arise when the impurities are connected to two conduction bands.

\subsection{Linear Chain Hamiltonian}
Following section II of \cite{kww1},
this transformation of the Hamiltonian can be divided into three main steps.
\begin{enumerate}
\item  One defines a new set of conduction-band operators $\{a_{k i \sigma}\}$, 
in terms of which the Hamiltonian takes a one-dimensional, continuum form. Each of the leads is assumed to consist of a single flat band, with density-of-states $\rho$ and bandwidth $2D$. The natural lead-dot hybridization parameter is then the quantity $\Gamma=\pi\rho V^2$, and the DQD Hamiltonian becomes

\begin{eqnarray}
\fl\frac{H}{D}=\sum_{i,\sigma} \left\{\int_{-1}^1 k \aadkisig\aakisig \rmd k + \left(\frac{\Gamma}{\pi D}\right)^{1/2}\int_{-1}^{1} (\aadkisig\cisig + \mathrm{h.c.})\rmd k\right\}\nonumber\\
\label{eq:hcont}
+\frac{1}{D}\sum_{i}(\epsilon\nhat_{i} + U\nhat_{i\uparrow}\nhat_{i\downarrow})+\frac{U'}{D}\nhat_{L}\nhat_{R}.
\end{eqnarray}

\item Both conduction bands are then divided symmetrically about $k=0$ into logarithmic intervals, the $n$th of which (for $k>0$) spans the range $\Lambda^{-(n+1)}<k<\Lambda^{-n}$. Only a single conduction-band state from each interval --- the fully symmetric linear combination --- couples to the impurity. All other conduction band states are 
neglected at this stage.

\item Finally, one performs a unitary transformation of the conduction-band states, 
to write each lead as a simple linear chain. The transformation is chosen such that only one site from each chain couples to the corresponding dot, by defining the annihilation operator for this site as
\begin{equation}
\label{eq:fnoughtdef}
\fpd_{0i\sigma}=\frac{1}{\sqrt{2}}\int_{-1}^1\rmd k ~\aakisig. 
\end{equation}
The remaining sites in each chain are described by operators $\fpd_{ni\sigma}$ (with somewhat more complicated expansions in terms of the $\aakisig$s, see \cite{kww1}), and the Hamiltonian thus takes the form
\begin{eqnarray}
\fl\frac{H}{D}=\sum_{i,\sigma}\left[\case{1}{2}(1+\Lambda^{-1})\sum_{n=0}^\infty \Lambda^{-n/2}\xi_n (\fd_{ni\sigma}\fpd_{n+1,i\sigma} + \mathrm{h.c.}) + \left(\frac{2\Gamma}{\pi D}\right)^{1/2}(\fd_{0i\sigma}\cisig+\mathrm{h.c.})\right]\nonumber\\
\label{eq:hdiscdef}
+\frac{1}{D}\sum_{i}(\epsilon\nhat_{i} + U\nhat_{i\uparrow}\nhat_{i\downarrow})+\frac{U'}{D}\nhat_{L}\nhat_{R}
\end{eqnarray}
where $\xi_n=(1-\Lambda^{-n-1})/[(1-\Lambda^{-2n-1})(1-\Lambda^{-2n-3})]^{1/2}$
(with $\xi_n \rightarrow 1$ for $n \gg 1$).
\end{enumerate}
\subsection{Iterative Diagonalization}
As originally discussed in \cite{wilson}, an accurate description of the ground state cannot be obtained  
 simply by diagonalising a truncated version of the linear-chain Hamiltonian (\ref{eq:hdiscdef}).
This is because the most important conduction band states at low-temperatures (those near the Fermi level) map onto the $\fpd_{ni\sigma}$ sites with the largest values of $n$. To determine the true nature of the ground state, it is necessary to retain \emph{all} conduction-band chain sites; this is achieved in practice by using an iterative procedure to diagonalize the Hamiltonian.

The first step is to write (\ref{eq:hdiscdef}) as
\begin{equation}
\label{eq:fullhlim}
H=\lim_{N\to\infty}\case{1}{2}(1+\Lambda^{-1})D\Lambda^{-(N-1)/2} H_N,
\end{equation}
where the $N$-site Hamiltonian $H_N$ is itself given by
\begin{eqnarray} 
\fl H_N=\Lambda^{(N-1)/2}\sum_{i,\sigma}\left[\sum_{n=0}^{N-1} \Lambda^{-n/2}\xi_n (\fd_{ni\sigma}\fpd_{n+1,i\sigma} + \mathrm{h.c.}) + \bar\Gamma^{1/2}(\fd_{0i\sigma}\cisig+\mathrm{h.c.})\right]\nonumber\\
\label{eq:hndef}
+\Lambda^{(N-1)/2)}\left[\sum_{i}(\bar\epsilon\nhat_{i} + \bar U\nhat_{i\uparrow}\nhat_{i\downarrow})+\bar U'\nhat_{L}\nhat_{R}\right],
\end{eqnarray}
with dimensionless coupling constants
\begin{eqnarray}
\bar \Gamma=\left(\frac{2}{1+\Lambda^{-1}}\right)^2\frac{2\Gamma}{\pi D}\qquad\bar\epsilon=\left(\frac{2}{1+\Lambda^{-1}}\right)\frac{\epsilon}{D}\nonumber\\
\bar U=\left(\frac{2}{1+\Lambda^{-1}}\right)\frac{U}{D}\qquad\bar U'=\left(\frac{2}{1+\Lambda^{-1}}\right)\frac{U'}{D}.
\end{eqnarray}
The key point here is that $H_N$ satisfies the recursion relation
\begin{equation}
\label{eq:recurdef}
\fl H_{N+1}=\Lambda^{1/2}H_N+\xi_N\sum_{i,\sigma}(\fd_{Ni\sigma}\fpd_{N+1,i\sigma}+\mathrm{h.c.})~~ \equiv ~ \mathcal T(H_N).
\end{equation}
Equation (\ref{eq:recurdef}) has the status of an RG transformation, and 
lies at the heart of the approach \cite{wilson,kww1,kww2}.
This recursion relation could in principle be implemented directly, by 
diagonalizing $H_{N+1}$ in the direct-product basis of the states of $H_N$ and the 16 states obtained by placing electrons in the $(N+1)$th orbitals of the two chains. In practice, however, the computation time required for the diagonalization of $H_{N+1}$ is greatly reduced by making use of its symmetries. As the coupling between the dots is purely capacitive, it is readily shown that $H_N$ commutes with the individual charge operators for the L/R subsystems
\begin{equation}
\label{eq:qnidef}
\hat Q_{Ni}=\sum_{n=0}^{N}\left(\sum_{\sigma}\fd_{ni\sigma}\fpd_{ni\sigma}-1\right)+\left(\sum_{\sigma}\cdisig\cisig-1\right)
\end{equation}
and also the left- and right-hand total spin operators
\begin{equation}
\label{snidef}  
\hat \bi{S}_{Ni}=\case{1}{2}\sum_{n=0}^{N}\sum_{\sigma,\sigma'}\fd_{ni\sigma}\bsigma_{\sigma\sigma'}\fpd_{ni\sigma'} + \case{1}{2}\sum_{\sigma,\sigma'}\cdisig\bsigma_{\sigma\sigma'}c^{\phantom\dagger}_{i\sigma'}
\end{equation}
where $\bsigma$ denotes the Pauli matrices. It is thus possible to label each eigenstate of $H_N$ using seven quantum numbers, i.e. $|Q_L,S_L,S_{Lz};Q_R,S_R,S_{Rz};r\rangle$; where $S_i$ and $S_{iz}$ represent respectively
the magnitude and $z$-component of spin on side $i$ ($=L/R$), and 
the index $r$ labels the different states of each charge and spin subspace.
By using the states of $H_N$ to form basis states for $H_{N+1}$ which are also eigenstates of $Q_{N+1,i}$, $S_{N+1,i}$ and $S_{z,N+1,i}$, it can be shown that the $H_{N+1}$ can be diagonalized \emph{independently} within each subspace, thereby reducing the computation time significantly. Furthermore, as the states within each spin multiplet are degenerate, one can use the Wigner-Eckart theorem \cite{kww1} to avoid having to consider each of them explicitly. It is then unnecessary to retain the $S_{iz}$ labels, which we thus omit from here onward. 

It is of course impossible to perform the iterative diagonalization exactly --- the dimensionality of the Hilbert space grows by a factor of 16 at each step. Even 
exploiting the symmetries just described, the complete numerical diagonalization of $H_N$ becomes unfeasible after just a few iterations. The solution \cite{wilson} is to retain only a certain number $N_\mathrm{s}$ of the lowest energy states of $H_N$ to calculate the states of $H_{N+1}$, with $N_\mathrm{s}$ chosen as a compromise between reducing the computation time while keeping the resulting truncation errors to a minimum. The results in this paper have typically been obtained using $N_{s}\sim 20,000$ (which we add excludes spin multiplicities). 
In addition, one must also take into account the error resulting from the logarithmic discretization of the Hamiltonian.
The discretized Hamiltonian (\ref{eq:hdiscdef}) is an exact transformation 
of (\ref{eq:hcont}) in the limit $\Lambda\to 1$. While one might naively
imagine that $\Lambda$ should be chosen as close to $1$ as possible, this is 
not so \cite{wilson,kww1} because the number of iterations needed to reach the 
low-energy conduction band states that dominate low-energy properties would 
become very large; 
not only increasing the calculation time, but more importantly allowing the 
cumulative effect of the truncation error from each iteration to become 
unacceptably large. It is in fact possible to make corrections for the 
discretization error resulting from even a relatively large $\Lambda$
\cite{wilson,kww1}; and for the present calculations we take $\Lambda = 3$.

\subsection{Thermodynamics}
\label{sec:calctherm}

  The temperature dependence of relevant thermodynamic functions can be calculated via the NRG as described in \cite{wilson,kww1}. In this paper, we give results for three `impurity' susceptibilities (borrowing the established terminology associated with Anderson-type impurity models). 
These are the spin, uniform charge and staggered charge susceptibilities, denoted by $\chi_s$, $\chi_c^+$ and $\chi_c^-$ respectively, and defined as 
\begin{eqnarray}
\label{eq:chisdef}
k_\mathrm{B}T\chi_s=(g\mu_\mathrm{B})^2\left\langle[\hat S_z-\langle\hat S_z\rangle]^2\right\rangle_\mathrm{imp}\\
\label{eq:chicpmdef}
k_\mathrm{B}T\chi_c^\pm=\case{1}{4}\left\langle[\hat N_L\pm\hat N_R-\langle\hat N_L\pm\hat N_R\rangle]^2\right\rangle_\mathrm{imp}
\end{eqnarray}
where $\hat S_z$ is the $z$-component of the total spin, and $\hat N_L$ ($\hat N_R$) is the total charge operator for the left (right) lead and dot subsystem 
(as usual, $\langle \hat{\Omega} \rangle_{\mathrm{imp}} = \langle \hat{\Omega} \rangle - \langle \hat{\Omega} \rangle_{0}$ with $\langle \hat{\Omega} \rangle_{0}$ 
denoting a thermal average in the absence of the impurities/dots). Note that the staggered charge susceptibility $\chi_c^-$ is effectively a `charge pseudospin' analogue of the spin susceptibility $\chi_{s}$, and that at the point $U'=U$ where
the model has $SU(4)$ symmetry (see (\ref{eq:hsufour})),
$\chi_c^- = \chi_s$ (as discussed further in \S 5,6).
  In addition, we shall use the impurity free energy
$F=-k_\mathrm{B}T\ln \left (Z/Z_0 \right )$
(with $Z_0$ the partition function in the absence of the two dots) to calculate the impurity entropy $ S = -\partial F/\partial T$,
and the linear specific heat coefficient $\gamma$ given by
$C(T)=-T\frac{\partial^2 F}{\partial T^2}\sim\gamma T+\mathcal{O}(T^3)$.
From here on, we will take $k_B=g\mu_B=1$ for convenience.
\section{Fixed points}
\label{sec:fixedpoints}
Before discussing the full numerical results, we examine the fixed points that control the NRG flows. 
In particular we shall see that there are two distinct classes of \emph{stable} fixed points, the mere existence of which implies a quantum phase transition in the model above a critical interdot interaction $U'_\mathrm{c}$.

Two initial remarks on the general nature of NRG fixed points should be made.
First, the energies of the Hamiltonian $H_N$ (\ref{eq:hndef}) are well known  \cite{wilson} to depend fundamentally on whether $N$ is even or odd. One works therefore with fixed points of $\mathcal T^2$, the fixed point Hamiltonian $H^*$ satisfying
$\mathcal T^2 (H^*) = H^*$.
The second, more subtle point is that each successive RG transformation increases 
the dimensionality of the Hilbert space of the problem; no Hamiltonian can thus
remain entirely unaffected by $\mathcal T^2$. But there do exist Hamiltonians, $H^*_N$, for which the \emph{low-lying} states (i.e.\ the $l$th lowest excitations, with $l\ll N$) remain asymptotically unaffected by the RG transformation; it is these which constitute the fixed points of the NRG procedure.

The set of NRG fixed points consistent with the symmetries of the DQD Hamiltonian 
(\ref{eq:hdiscdef}) may be deduced by setting each of its bare parameters to a value of either zero or infinity. This
leaves in all cases a pair of `free' conduction chains, plus a piece that describes any remaining dot degrees-of-freedom. Since the free conduction chain energy levels themselves converge rapidly \cite{wilson} with increasing $N$, while the isolated dot degrees-of-freedom are entirely unaffected by the NRG transformation, this decoupled Hamiltonian is necessarily a fixed point of $\mathcal T^2$.

\Tref{tab:fixedpoints} lists the five distinct fixed points resulting from the procedure sketched above.
\begin{table}
\caption{\label{tab:fixedpoints}Fixed points of the DQD system, obtained from (\ref{eq:hndef}) \\ 
by setting  $\Gamma$, $U$ and $U'$ to values of either 0 or $\infty$.}
\begin{indented}
\item[]\begin{tabular}{@{}lllll}
\br
Name&Abbreviation&$\Gamma$&$U$&$U'$\\
\mr
Free Orbital&\fo&0&0&0\\
$SU(2)$ Local Moment&\lmtwo{}&0&$\infty$&0\\
$SU(4)$ Local Moment&\lmfour{}&0&$\infty$&$\infty$\\
Charge Ordered&\co&0&0&$\infty$\\
Strong Coupling&\stc&$\infty$&0&0\\
\br
\end{tabular}
\end{indented}
\end{table}
The three fixed point Hamiltonians with $U'=0$ 
are obviously just `doubled' versions of the corresponding Hamiltonians for the symmetric Anderson model \cite{kww1} (although the physical behaviour near these fixed points will generally differ from that of the single-impurity case because  corrections to the fixed points can include terms that couple the two channels). The other two fixed points, \lmfour{} and \co, have no analogues in the AIM, and thus generate the distinct physics of the DQD model.

The stability of each fixed point in \tref{tab:fixedpoints} can be investigated by linearising the NRG transformation in its vicinity.
Any perturbation $\delta H$ of the fixed point that is consistent with its underlying symmetries, is known \cite{wilson} to have the following expansion in terms of the eigenvectors $O_l^*$, and corresponding eigenvalues $\lambda_l^*$, of the matrix $\mathcal{L}^*$ that describes the transformation:
\begin{equation}
\label{eq:fpexpand}
\Lambda^{(N-1)/2}\delta H=\sum_l C_l\lambda_l^{*N/2}O_l^*,
\end{equation}
(with $C_l$ a constant independent of $N$). 
By analysing the possible $\delta H$s in (\ref{eq:fpexpand}), one can deduce the eigenvalues $\lambda_l^*$ by equating the $N$-dependences of both sides. If any of these $\delta H$s generate relevant eigenvectors, the fixed point is unstable.

For each fixed point of the DQD system, \tref{tab:fixedpointcorrs} shows the resultant perturbations that generate the most relevant eigenvectors of the linearised NRG transformation. Most are straightforward to analyse in practice. The free orbital fixed point is readily shown to be unstable, and the strong coupling fixed point to be stable. Each of the local moment fixed points is marginal to leading (linear) order, but our numerical results in \sref{sec:numrestherm} show they are both unstable under the full NRG transformation. The charge ordered (CO) fixed point by contrast is more subtle; we thus devote the next section to consideration of its behaviour.

\begin{table}
\caption{\label{tab:fixedpointcorrs}Leading corrections to the fixed points of the DQD. The quantity $\btau_i$ represents a spin-$\half$ on dot $i$, while $\alpha$ is an anomalous exponent discussed in the text. In this table, we adopt the convention that repeated spin indices $\sigma$ are to be implicitly summed over unless otherwise stated.}
\begin{tabular}{@{}lllll}
\br
Fixed point&Label&Perturbation&Eigenvalue&Stability\\
\mr  

\fo&$\delta H_1$&$\sum_{i} (\cdisig\fpd_{0i\sigma}+\mathrm{h.c.})$&$\Lambda^{1/2}$&Unstable\\
&$\delta H_2$&$\sum_{i} (\cdisig\cisig-1)^2$&$\Lambda$\\
&$\delta H_3$&$\hat n_{L}\hat n_{R}$&$\Lambda$\\
\lmtwo{}&$\delta H_4$&$\sum_{i}(\fd_{0i\sigma}\bsigma_{\sigma,\sigma'}\fpd_{0i\sigma'})\cdot\btau_i$&$1$&Marginally-unstable\\
\lmfour{}&$\delta H_5$&$\sum_{i,j}c^\dagger_{i\sigma}c^{\phantom\dagger}_{j\sigma'}\fd_{0j\sigma'}\fpd_{0i\sigma}$&$1$&Marginally-unstable\\
\stc&$\delta H_6$&$\sum_{i}(\fd_{1i\sigma}\fpd_{2i\sigma}+\mathrm{h.c.})$&$\Lambda^{-1}$&Stable\\
&$\delta H_7$&$\sum_{i}(\fd_{1i\sigma}\fpd_{1i\sigma}-1)^2$&$\Lambda^{-1}$\\
&$\delta H_8$&$\prod_i(\fd_{1i\sigma}\fpd_{1i\sigma}-1)$&$\Lambda^{-1}$\\
\co&$\delta H_9$&$\sum_{i}(\nhat_i-1)\fd_{0i\sigma}\fpd_{0i\sigma}$&$1$&Stable if $\alpha>0$\\
&$\delta H_{10}$&$\prod_\sigma (c^\dagger_{R\sigma}\fpd_{0R\sigma}\fd_{0L\sigma}c^{\phantom\dagger}_{L\sigma}+\mathrm{h.c.})$&$\Lambda^{-2\alpha}$
\\
\br
\end{tabular}
\end{table}

\subsection{The charge ordered fixed point}
\label{sec:cofixedpoint}
The CO fixed point Hamiltonian can be obtained by setting $U'=\infty$ in (\ref{eq:hndef}). This decouples the dots from the leads, leaving a pair of free conduction bands plus the two degenerate dot charge configurations $(n_L,n_R)= (0,2)$ and $(2,0)$. To avoid excessive use of projection operators in the following, it is implicit that the Hilbert space in which the CO fixed point Hamiltonian acts contains these 
dot configurations only.

As seen in \tref{tab:fixedpointcorrs}, two corrections to the fixed point govern its stability. We begin by discussing the marginal term 
\begin{equation}
\label{eq:hncops}
\fl\delta H_9=\sum_{i,\sigma}(\nhat_i-1)\fd_{0i\sigma}\fpd_{0i\sigma} \equiv \half(\nhat_L-\nhat_R)\sum_\sigma\Bigl(\fd_{0L\sigma}\fpd_{0L\sigma} - \fd_{0R\sigma}\fpd_{0R\sigma}\Bigr),
\end{equation}
which describes potential scattering of the conduction electrons. The key point here is that although the magnitude of this potential scattering is equal for both the $L$ and $R$ conduction bands, the signs are opposite and \emph{correlated} with the two charge configurations of the dots. It is this correlation 
that is responsible for much of the interesting behaviour of the DQD model.

As shown originally in \cite{kww2}, the effect of potential scattering on the conduction band chains can be conveniently incorporated into the definition of the fixed point itself. Following the notation of \cite{kww2} we introduce a coupling constant $\tilde K \geq 0$ to specify its magnitude---such that the appropriate picture is of a \emph{line} of CO fixed points with different values of $\tilde K$---and thus obtain the following CO fixed point Hamiltonian: 
\begin{eqnarray}
\label{eq:hncok}
\fl H_{N,\mathrm{CO}}^*(\tilde K)=\Lambda^{(N-1)/2}\sum_{i,\sigma}\Biggl[\sum_{n=0}^{N-1}\Lambda^{-n/2}\xi_n(\fd_{ni\sigma}\fpd_{n+1,i\sigma}+\mathrm{h.c.})
+\tilde K(\nhat_i-1)\fd_{0i\sigma}\fpd_{0i\sigma}\Biggr].
\end{eqnarray}
In the Appendix, we point out that a continuum version of this Hamiltonian is precisely the `full' low-energy model for the DQD system when $U'\gg U$;
and by comparison of the discrete and continuum forms obtain a relationship between 
$\tilde K$ and the bare parameters of the model, which we show to be in excellent agreement with $\tilde K$ obtained from the full NRG calculations at large $U'$. As far as the present discussion is concerned, the most important result to be drawn from the continuum version of (\ref{eq:hncok})
is that the potential scattering generates equal and opposite phase shifts in the two leads, the magnitude of which can be related to $\tilde K$ via 
\begin{equation}
\label{eq:cophaseshift}
\delta=\tan^{-1}\left[\frac{\pi(1-\Lambda^{-1})\tilde K}{2\ln\Lambda}\right],
\end{equation}
which result will shortly prove useful.

Given that $\delta H_9$ above is exactly marginal, the stability of each CO fixed point is found to be controlled by the term $\delta H_{10}$ in \tref{tab:fixedpointcorrs}, namely $\delta H_{10}=A+A^\dagger$ with
\begin{equation}
\label{eq:deltahten}
A=\prod_\sigma   
c^\dagger_{R\sigma}\fpd_{0R\sigma}\fd_{0L\sigma}c^{\phantom\dagger}_{L\sigma}.
\end{equation}
This operator describes four-electron cotunnelling processes that interconvert the $(0,2)$ and $(2,0)$ dot configurations, and thereby switch the signs of the potential scattering on each lead. The relevance of $\delta H_{10}$ is not obvious from simple power counting \cite{garst,perakis}; the action of \eref{eq:deltahten} on the eigenstates of $H_{N,\mathrm{CO}}^*(\tilde K)$ is very similar to the non-perturbative X-ray edge problem \cite{mahan, nozdedom} and hence requires a somewhat more sophisticated treatment such as bosonization \cite{schotte}. By considering the long-time behaviour of the correlation function $\langle A^\dagger(t)A\rangle_0$, one can show that the eigenvalue associated with $\delta H_{10}$ goes as $\Lambda^{-2\alpha}$, where $\alpha$ is an anomalous exponent that is related to the conduction band phase shift $\delta$ via 
\begin{equation}
\label{eq:alphadef}
\alpha=-\frac{1}{2} + \left(\frac{2\delta}{\pi}-1\right)^2.
\end{equation}
This anomalous exponent arises also for a particular limit of the Ising-coupled 
Kondo impurity model \cite{garst}, a point which to which we return in \sref{sec:effhckondo}.

The stability of each CO fixed point depends on the corresponding sign of $\alpha$, which can be calculated directly from \eref{eq:cophaseshift} and \eref{eq:alphadef}. Since $\tilde K \in (0,\infty)$ it is easy to see that the line of CO fixed points divides at a critical $\tilde K_\mathrm{c}$ into two classes: stable for $\tilde K < \tilde K_\mathrm{c}$ (i.e. $\alpha>0$), and unstable for $\tilde K > \tilde K_\mathrm{c}$ ($\alpha<0$). The critical point, at which $\alpha$ vanishes, thus corresponds to a phase shift $\delta_\mathrm{c}$ given by
\begin{equation}
\delta_\mathrm{c}=\frac{\pi}{2}\left(1-\frac{1}{\sqrt{2}}\right)
\end{equation} 
and hence a critical potential scattering strength 
\begin{equation}
\tilde K_c=\frac{2\ln\Lambda}{\pi(1-\Lambda^{-1})}\tan\left[\frac{\pi}{2}\left(1-\frac{1}{\sqrt{2}}\right)\right]
\end{equation}
(as we also confirm numerically in the Appendix).
Note that in the particle-hole symmetric limit considered explicitly, 
$\delta_\mathrm{c}$ is entirely independent of the bare material parameters. 
That is not however true in general --- deviation from particle-hole symmetry  generates additional potential scattering of the \emph{same} sign on the two leads. This breaks the simple `equal and opposite' relationship between the $L$ and $R$ phase shifts, and leads to a more complicated form for the anomalous exponent $\alpha$ (but as mentioned above does not alter the essential physics, which is wholly robust to departure from \emph{p-h} symmetry).

\subsection{Quantum phase transition and schematic flow diagram} 
Our analysis of the NRG fixed points shows two qualitatively distinct classes of stable fixed points, SC and CO. This of course implies the existence of a quantum phase transition in the system. Points in phase space that flow to the SC fixed point under renormalization comprise the `strong coupling' phase, where all the degrees of freedom associated with the dots are frozen-out at low temperatures, while the set of points which flow to the CO fixed point define a `charge ordered' phase which has a doubly-degenerate ground state. Although the phase transition can of course be approached in any direction, we will tend to envisage crossing from SC to CO by increasing $U'$ at a fixed $U$ and $\Gamma$; the transition then takes place at the critical point $U'=U'_c$.

\begin{figure}
\begin{center}
\includegraphics[height=7cm]{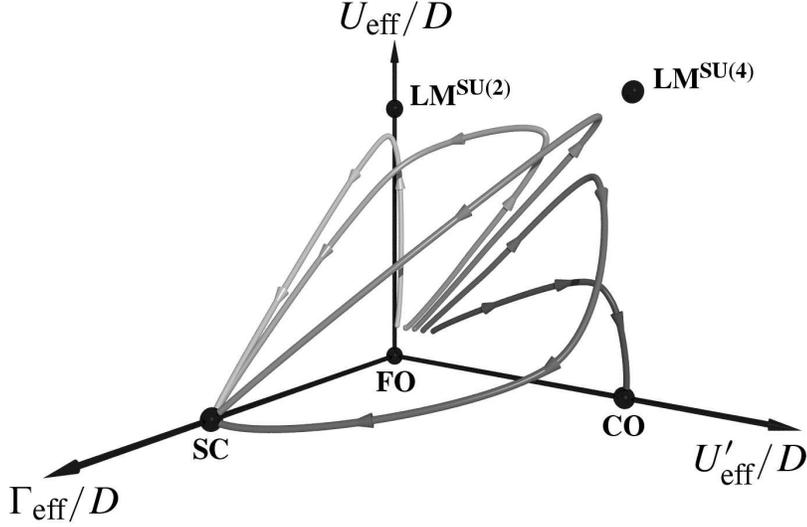}
\end{center}
\caption{\label{fig:flow}Schematic flow diagram for the DQD system. We consider the case when the bare $U\gg\Gamma$, but with $U$, $U'$ and $\Gamma$ all much smaller than the bandwidth $D$. The diagram shows the effect of increasing $U'$ at fixed $U$ and $\Gamma$ (light to dark flow lines); at a critical $U'_c$ the flow tends ultimately to CO instead of SC. (Note that in addition to the three effective parameters $U_\mathrm{eff}$, $U'_\mathrm{eff}$ and $\Gamma_\mathrm{eff}$ shown, there is an effective potential scattering $\tilde K$ generated in the vicinity of the CO fixed point.)}
\end{figure}

It is helpful at this stage to draw a schematic flow diagram showing how the various fixed points of the model are approached under different choices of the bare parameters $U$, $U'$ and $\Gamma$; this will be 
useful in the next section when we analyse how the various unstable fixed points affect the SC phase, and how the CO fixed point in particular controls the quantum phase transition. [The structure of the flow diagram is of course deduced from the full NRG calculations, as detailed in section 6.]
In \Fref{fig:flow} the flow diagram is drawn in the three-dimensional ($\Gamma_\mathrm{eff}/D,U_\mathrm{eff}/D,U'_\mathrm{eff}/D$) effective parameter space (we omit the potential scattering generated in the region of the CO fixed points, and thus represent schematically the line of CO fixed points by a single point). In this work we are primarily concerned with the strongly-correlated regime arising for $U/\Gamma\gg 1$, and hence take the bandwidth $D$ to be the largest energy scale; all flows thus begin near the FO fixed point, as shown in the figure.

When the bare $U'=0$, the flow is first to the \lmtwo{} fixed point, and then to the stable SC \cite{kww1}. Switching on a finite $U'$ brings the additional fixed points of the DQD system into play, as shown by the successively darker flow lines in \Fref{fig:flow}. When $U'=U$ (i.e.\ the `middle' flow line in \fref{fig:flow}), the NRG flow remains in the $U_\mathrm{eff}=U'_\mathrm{eff}$ plane throughout, heading to the \lmfour{} fixed point before ultimately the stable SC. With any deviation from this $SU(4)$ symmetry, the flows are drawn away from the the $U_\mathrm{eff}=U'_\mathrm{eff}$ plane, crossing over instead to \lmtwo{} (for $U'<U$) or CO (for $U'>U$). The physics around the CO fixed point is in fact crucial to the problem: we show in \sref{sec:phase} that there is a range of $U'$ above $U$ where the CO fixed point is unstable, the flows ultimately turning to SC at large $N$ as depicted by the fourth flow in \fref{fig:flow}. At a critical $U'_c$ however, the CO fixed point becomes stable and the quantum phase transition from the SC phase to the CO phase ensues. In \sref{sec:numrestherm} we analyse the numerical results to confirm the above picture, and identify the relevant energy scales at which the crossovers between the various fixed points occur. 
\section{Fixed point analytics}
\label{sec:fpa}
First, we focus on results that can be obtained analytically by considering how the leading corrections to the SC and CO fixed points affect the flows in their vicinities. In particular we show that the effective low-energy Hamiltonian of the system near criticality can be obtained from flows close to the CO fixed point, and analyse the behaviour near SC to obtain the leading $T\to 0$ behaviour of the SC phase.

\subsection{Effective Hamiltonian near the phase boundary}
\label{sec:effhckondo}
We begin with the CO fixed point. 
For $U'\approx U'_c$, there exists a range of $N$ over which the NRG Hamiltonians $H_N$ flow very close to the line of CO fixed points (see \fref{fig:flow}) and can therefore be approximated by $H_{N,CO}^*$ plus its leading corrections. To make this statement quantitative, suppose that the flow is close to the CO fixed points when $N=N_1$ and use (\ref{eq:hncok}) and (\ref{eq:deltahten}) to write the corresponding Hamiltonian as 
\begin{eqnarray}
\fl H_{N_1, CO}&\approx&H^*_{N_1,\co}(\tilde K)+\omega_{10}\Lambda^{-(N_1-1)/2}\delta H_{10}\nonumber\\
\fl &=&\Lambda^{(N_1-1)/2}\Biggl[\sum_{i,\sigma}\Biggl(\sum_{n=0}^{N_1-1}\Lambda^{-n/2}\xi_n(\fd_{ni\sigma}\fpd_{n+1,i\sigma}+\mathrm{h.c})+\tilde K(\nhat_i-1)\fd_{0i\sigma}\fpd_{0i\sigma}\Biggr)\nonumber\\
\label{eq:hnone}
\fl &&+\omega_{10}\prod_\sigma (c^\dagger_{R\sigma}\fpd_{0R\sigma}\fd_{0L\sigma}c^
{\phantom\dagger}_{L\sigma}+\mathrm{h.c.})\Biggr].
\end{eqnarray}
(with the constants $\tilde K$ and $\omega_{10}$ naturally determined by fitting the energy levels of (\ref{eq:hnone}) to those obtained from the full NRG calculations). 

Suppose that we are only interested in the low-temperature properties of the system (where by `low-temperature' we mean $T<T_{N_1} \sim D\Lambda^{-N_{1}/2}$).
These are calculated from the sequence of Hamiltonians $H_{N_1}$, $H_{N_1+1}$, $\ldots$, obtained by applying successive RG transformations (\ref{eq:recurdef}) to $H_{N_1}$. 
While we have here obtained $H_{N_1}$ from our analysis of the `full' DQD Hamiltonian (\ref{eq:hnought},{\it b}), it is clear that \emph{any} bare Hamiltonian that leads to the same $H_{N_1}$ under RG flow will show precisely the same thermodynamics for $T<T_{N_1}$; such a Hamiltonian thus constitutes an effective low-temperature model of the system. The simplest bare Hamiltonian that generates \eref{eq:hnone} is of the form
\begin{eqnarray}
\fl H=
\sum_{i,\bk,\sigma}\epsilon_\bk a^\dagger_{\bk i\sigma}a^{\phantom\dagger}_{\bk i\sigma}+ J_\perp\left(T^- \prod_{\sigma}\sum_{\bk,\bk'}a^{\phantom\dagger}_
{\bk R\sigma}a^\dagger_{\bk'L\sigma}+\mathrm{h.c.}\right)\nonumber\\
\label{eq:hncoeff}
+J_z T_z\sum_{\sigma,\bk,\bk'}(a^{\dagger}_{\bk L\sigma}a^{\phantom\dagger}_{\bk' L\sigma}-a^{\dagger}_{\bk R\sigma}a^{\phantom\dagger}_{\bk' R\sigma}),
\end{eqnarray}
where $T$ represents a spin-$\case{1}{2}$ pseudospin defined by the operators $T^+=\prod_\sigma c^\dagger_{L\sigma}c^{\phantom\dagger}_{R\sigma}$, $T^-=\prod_\sigma c^\dagger_{R\sigma}c^{\phantom\dagger}_{L\sigma}$ and $T_z=(\hat n_L-\hat n_R)/4$; and the coupling constants are $J_\perp=\case{1}{2}(1+\Lambda^{-1})\omega_{10}D$ and $J_z=(1+\Lambda^{-1})\tilde K D$.
The reader can verify that (\ref{eq:hncoeff}) gives the same $H_{N_1}$ as (\ref{eq:hnone}) by substituting it for the full DQD Hamiltonian in (\ref{eq:hnought},{\it b}) and repeating the subsequent steps of \sref{sec:nrg}. A new $H_N$ is obtained in place of (\ref{eq:hndef}), which is identical to (\ref{eq:hnone}) when $N=N_1$. 
Equation \eref{eq:hnone} thus captures the essential low-temperature degrees-of-freedom of the system close to the quantum phase transition.

It is 
however instructive to recast (\ref{eq:hncoeff})
in a somewhat different form. As it stands, the pseudospin flips described by $T^\pm$ occur in combination with electrons hopping between the left and right conduction bands. It is expedient to interchange the spin and $L/R$ labels ($\uparrow\leftrightarrow L$, $\downarrow\leftrightarrow R$) so that in this relabelled form the pseudospin flips are accompanied by spin flips of the conduction electrons. 
With this, (\ref{eq:hncoeff}) reduces to
\begin{eqnarray}
\label{eq:heffpb}
\fl H=\sum_{i,\bk,\sigma}\epsilon_\bk a^\dagger_{\bk i\sigma}a^{\phantom\dagger}_{\bk i\sigma}+J_\perp\left(T^- \prod_{i}\sum_{\bk,\bk'}a^{\phantom\dagger}_{\bk i\downarrow}a^\dagger_{\bk'i\uparrow}+\mathrm{h.c.}\right)\nonumber\\
+J_z T_z\sum_{i,\bk,\bk'}(a^{\dagger}_{\bk i\uparrow}a^{\phantom\dagger}_{\bk' i\uparrow}-a^{\dagger}_{\bk i\downarrow}a^{\phantom\dagger}_{\bk' i\downarrow}).
\end{eqnarray}
This effective low-energy Hamiltonian is generically valid as
$\tilde U'\to \tilde U'_c$. A Hamiltonian of this form has been investigated previously, in the problem of a pair of Ising-coupled Kondo impurities 
\cite{garst}. The latter model displays a quantum phase transition of Kosterlitz-Thouless \cite{kosterlitz} (KT) type, and in \sref{sec:phase} we show that the 
phase transition in our DQD system is also of KT type.

\subsection{Thermodynamics of the strong coupling phase}
\label{sec:thermosc}
We turn now to the leading $T\to 0$ behaviour of the SC phase, which can be obtained from the SC fixed point Hamiltonian $H_{N,SC}^*$ and its most relevant corrections given in \tref{tab:fixedpointcorrs}. We write the effective NRG Hamiltonian as
\begin{equation}
\label{eq:hnsccorr}
\fl H_{N,SC}\approx H_{N,SC}^*+\omega_6\Lambda^{-(N-1)/2}\delta H_6+\omega_7\Lambda^{-(N-1)/2}\delta H_7+\omega_8\Lambda^{-(N-1)/2}\delta H_8,
\end{equation}
(with the coupling constants $\omega_6$, $\omega_7$ and $\omega_8$ determined 
numerically). It is straightforward but lengthy to calculate the $T\to 0$ thermodynamics of this effective Hamiltonian; one simply uses the results of \sref{sec:calctherm}, treating the fixed point corrections to leading order in perturbation theory (see e.g.\ section V of \cite{kww1}). 
The final results for the impurity susceptibilities and linear specific heat coefficient as $\Lambda\to 1$, are found to be
\begin{eqnarray}
\label{eq:chicpmsc}
\chi_{c}^\pm=\frac{8}{D}\frac{2}{1+\Lambda^{-1}}\frac{\alpha_0\alpha_1}{\ln\Lambda}\left[-\omega_6-(\omega_7\pm\omega_8)\frac{\alpha_0^3}{\alpha_1\ln\Lambda}\right]\\
\chi_s=(g\mub)^2\frac{2}{D}\frac{2}{1+\Lambda^{-1}}\frac{\alpha_0\alpha_1}{\ln\Lambda}\left[-\omega_6+\omega_7\frac{\alpha_0^3}{\alpha_1\ln\Lambda}\right]\\
\label{eq:gamsc}
\gamma=(\pi k_B)^2\frac{8}{3D}\frac{2}{1+\Lambda^{-1}}\frac{\alpha_0\alpha_1}{\ln\Lambda}\left[-\omega_6\right],
\end{eqnarray}
where $\alpha_0$ and $\alpha_1$ are dependent on the discretization parameter $\Lambda$, viz
\begin{equation}
\alpha_0=\left[\case{1}{2}(1-\Lambda^{-1})\right]^{\case{1}{2}}\qquad\qquad\alpha_1=\left[\case{1}{2}(1-\Lambda^{-3})\right]^{\case{1}{2}}.
\end{equation}

Equations (\ref{eq:chicpmsc})--(\ref{eq:gamsc}) provide a convenient means of calculating accurately the low-temperature thermodynamics from the NRG energy levels obtained numerically. More importantly however, their dependences on the three parameters $\omega_6$, $\omega_7$ and $\omega_8$ enable us to deduce some exact results. We shall express these in the language of Wilson ratios, defined by
\begin{equation}
\label{eq:wilsondef}
R_s=\frac{4}{3}\frac{(\pi k_B)^2}{(g\mub)^2}\frac{\chi_s}{\gamma}\qquad\qquad R_c^\pm=\frac{4(\pi k_B)^2}{3}\frac{\chi_c^\pm}{\gamma}
\end{equation}
where we add for later use that $T_K = 1/\gamma$ provides a suitable definition of the Kondo scale $T_K$ characteristic of the SC phase.
Here $R_s$ is the double dot analogue of the usual `spin' Wilson ratio, and $R_c^\pm$ make up a pair of `charge' Wilson ratios. It is straightforward to show from (\ref{eq:chicpmsc})--(\ref{eq:gamsc}) that these Wilson ratios can be expressed in terms of the coupling constants of the effective Hamiltonian (\ref{eq:hnsccorr}) as
\begin{equation}
\label{eq:wilsons}
R_s=1-\frac{\omega_7}{\omega_6}\frac{\alpha_0^3}{\alpha_1\ln\Lambda}.
\qquad\qquad
R_c^\pm=1+\frac{(\omega_7\pm\omega_8)}{\omega_6}\frac{\alpha_0^3}{\alpha_1\ln\Lambda},
\end{equation}
and therefore that the three Wilson ratios are related by
\begin{equation}
\label{eq:wilsonrel}
R_c^+ + R_c^- + 2R_s=4.
\end{equation}
This $\Lambda$-independent result is satisfied throughout the SC phase, and we believe it to be exact.
We can moreover go further by focusing on three high-symmetry limits of the model,  where it is possible to determine the precise values of the Wilson ratios. We end this section by considering these limits in turn.

\subsubsection{The $SU(2)$ limit.} When $U'=0$, the left and right channels of the system are disconnected and it is clear that the coupling constant $\omega_8$ of the effective model (\ref{eq:hnsccorr}) vanishes.
We then of course recover from (\ref{eq:wilsons}) and (\ref{eq:wilsonrel}) the well-known results for the AIM \cite{hewson}: the two charge Wilson ratios coincide ($R_c^+=R_c^-\equiv R_c$), and $R_s+R_c=2$ for all $U$. In the strong coupling limit, 
$\tilde{U} \gg 1$, the system maps onto a Kondo model where each dot is strictly singly-occupied; the charge Wilson ratio $R_c$ then vanishes and $R_s=2$ \cite{wilson,nozieres}.
\subsubsection{The $SU(4)$ limit.} 
At the point $U'=U$, 
we can make use of the underlying $SU(4)$ symmetry of the Hamiltonian (see (\ref{eq:hsufour})); for (\ref{eq:hnsccorr}) to be likewise $SU(4)$-invariant, it is  necessary and sufficient that $\omega_8=2\omega_7$, such that
\begin{eqnarray}
H_{N,SC}\approx H_{N,SC}^*+\omega_6\Lambda^{-(N-1)/2}\sum_{m}(\fd_{1m}\fpd_{2m}+\mathrm{h.c.})\nonumber\\
+\omega_7\Lambda^{-(N-1)/2}
[\sum_m\fd_{1m}\fpd_{1m}-2
~]^{2}
\end{eqnarray}
with 
$m \equiv (i,\sigma)$ as before. Using $\omega_8=2\omega_7$ in (\ref{eq:wilsons}) gives $R_s=R_c^-$ for all $U'=U$ in the SC phase. 
In the strong-coupling limit $\tilde{U} \gg 1$ in particular, 
the total charge on the dots is strictly $n=2$, $R_c^+$ vanishes again and it follows from (\ref{eq:wilsonrel}) that \cite{deleo}
\begin{equation}
\label{eq:doodahday}
R_s=R_c^-=\case{4}{3}, ~~ R_\mathrm{c}^+=0\qquad\mbox{when}\qquad \tilde{U}'=\tilde{U}\to\infty .
\end{equation}

\subsubsection{The $U'\to U'^{-}_{c}$ limit.}
In \sref{sec:effhckondo}, we showed that the low-temperature ($T\ll U'-U$) properties of the system near the phase transition can be obtained from a Hamiltonian of the form (\ref{eq:hnone}). Here we can use one of the symmetries of this Hamiltonian to deduce a relation between $\omega_7$ and $\omega_8$ when $U'$ is close to $U'_c$ 
in the SC phase.
As a result of its reduced dot degrees-of-freedom compared to the full DQD Hamiltonian (\ref{eq:hndef}), the effective Hamiltonian (\ref{eq:hnone}) is invariant not only under a complete particle-hole transformation, but also under a combined particle-hole transformation and interchange of the $L$ and $R$ labels \emph{in a single spin channel}. Such a transformation is performed as follows for the $\uparrow$-spin channel
\begin{equation}
c^{\dagger}_{L/R\uparrow}\leftrightarrow c^{\phantom\dagger}_{R/L\uparrow}\qquad\mbox{and}\qquad\fd_{nL/R\uparrow}\leftrightarrow(-1)^n\fpd_{nR/L\uparrow},
\end{equation}
with the $\downarrow$-spin operators left unchanged. It is straightforward to show that for $H_{N,SC}$ to be similarly invariant under this transformation, $\omega_8=-2\omega_7$, and thus from (\ref{eq:wilsons}) that $R_s=R_c^+$ as $U'\to U'^{-}_c$. 
Further, as shown in section 6 the Kondo scale $T_K$ vanishes as
the transition is approached (and thus $\gamma$ diverges), while $\chi_s$ remains
finite; hence 
\begin{equation}
\label{eq:doodah}
R_s=R_c^+=0, ~~ R_\mathrm{c}^-=4\qquad\mbox{when}\qquad U'\to U_{c}'^{-}
\end{equation}
(holding for any $U$).
\section{Numerical results: thermodynamics}
\label{sec:numrestherm}
We now present numerical results obtained from the NRG calculations. While much can been deduced from the general RG framework itself, as above, the numerics provide  accurate and detailed results for thermodynamics and dynamics over a wide range of temperature and frequency scales. 
We first determine the phase boundary between the SC to CO phases in the $(U,U')$ plane, before analysing the Kondo
scale that controls the low-energy behaviour of the SC phase and vanishes as the
SC$\rightarrow$CO transition is approached.
Typical temperature dependences of the `impurity' entropy, spin- and staggered charge-susceptibilities with increasing $U'$ are then considered, and shown to exhibit universal scaling behaviour in the SC phase.

\subsection{Phase diagram and the Kondo scale}
\label{sec:phase}
\begin{figure}
\centering\includegraphics{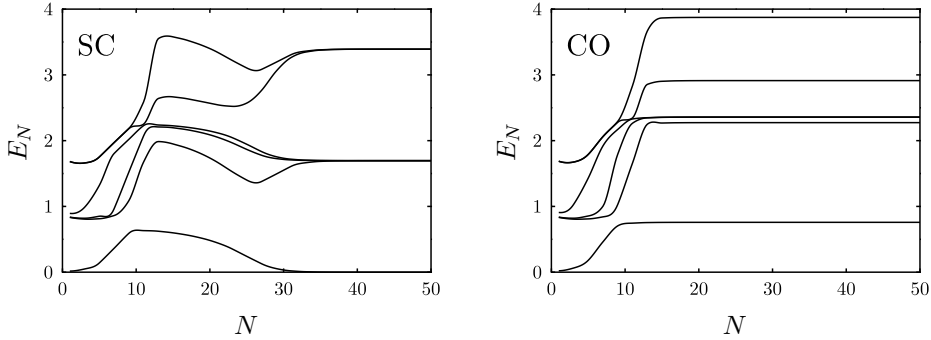}
\caption{\label{fig:energyflows} Lowest six energy levels of the $(Q_L,S_L,Q_R,S_R)=(1,0,0,\case{1}{2})$ subspace of $H_N$, as a function of $N$, at characteristic points deep in the SC ($\tilde U=7$, $\tilde U'=6$) and CO ($\tilde U=7$, $\tilde U'=8$) phases. }
\end{figure}

We determine the phase boundary in the $(U,U')$ plane by a simple bisection method, classifying points therein as belonging either to the SC or CO phase by analysing their corresponding NRG flows.
 For a particular choice of $\tilde{U} = U/(\pi \Gamma)$ and $\tilde{U}' = (U'/\pi \Gamma)$, the fixed point ultimately reached as $N\to \infty$ can be identified by comparing the energy levels of the full NRG calculation  with those easily obtained from the single-particle energies of the fixed point Hamiltonians (\ref{eq:hnsccorr}) and (\ref{eq:hncok}). The energy level patterns that arise in the two phases are quite distinct, so the identification of the ground state is  straightforward. To illustrate this, figure \ref{fig:energyflows} shows the $N$-dependence of the six lowest energy levels in the $(Q_L,S_L,Q_R,S_R)=(1,0,0,\case{1}{2})$ subspace of $H_N$, for two characteristic points `deep' in the SC and CO phases: $\tilde U=U/(\pi\Gamma)=7$, with $\tilde U'=U'/(\pi\Gamma)=6$ 
 (SC) and $\tilde U'=8$ (CO). The initial flows in both phases are clearly very similar, as expected because both flows begin near the FO fixed point. With increasing $N$ however, the SC and CO energy levels begin to depart from each other, due to the influences of the different fixed points on the two flows. For large $N$, the SC and CO phase levels converge to qualitatively different limits. The pattern of levels shown for the SC phase in \fref{fig:energyflows} is in fact the same for all points in the SC phase, while in the CO phase the values of the fixed point energies depend on the potential scattering $\tilde K$ (see (\ref{eq:hncok})), which varies with the choice of bare model parameters. 

\begin{figure}
\centering\includegraphics{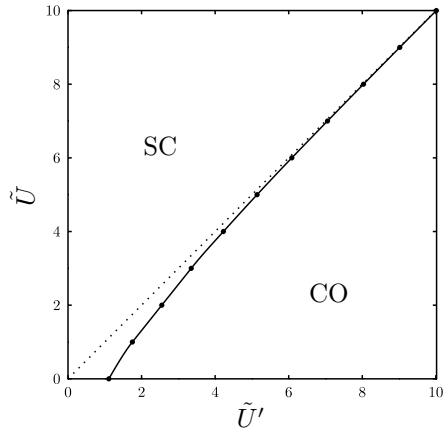}
\caption{\label{fig:pb} Phase diagram in the $(\tilde U',\tilde U)$ plane. The solid line is the phase boundary between the SC and CO phases, and the dotted line is the $SU(4)$ line $\tilde U'=\tilde U$.}
\end{figure}

Figure \ref{fig:pb} shows the resultant phase boundary.
As a function of $\tilde{U}$, the critical $\tilde{U}'$ is finite for all $\tilde{U}\ge 0$ (and moreover exceeds $\tilde U$, the boundary between $(1,1)$ and $(2,0)/(0,2)$ configurations in the atomic limit). In strong-coupling in particular ($\tilde{U} \gg 1$), we find $U'_c-U$ to be exponentially small, taking the form $(U'_c/U-1)\sim c\tilde U^{1/4}\exp(-\pi^2 \tilde U/16)$. We show below that extension of the SC phase beyond the atomic limit boundary reflects the formation of a \emph{charge-Kondo} state, where the doubly-degenerate charge pseudospin that exists freely in the CO phase is quenched by coupling to the leads. Simple physical arguments~\cite{ddprl} (reprised in the Appendix) show that the stability of this charge-Kondo state depends on the magnitide of $\tilde U'-\tilde U$ relative to the $SU(4)$ Kondo temperature, and hence we now turn to consideration of $T_K$ in the SC phase.

\begin{figure}
\centering\includegraphics{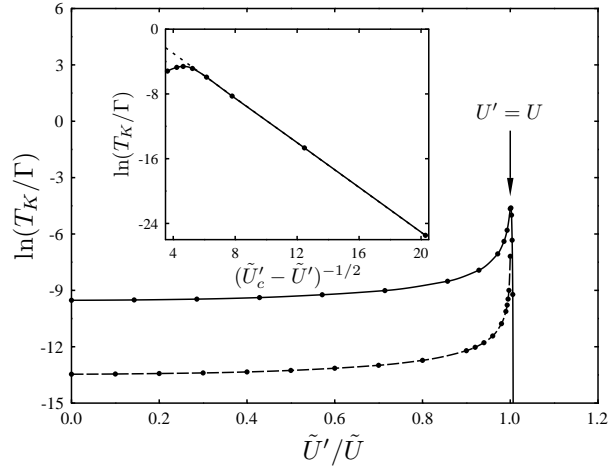}
\caption{\label{fig:kondoscale}Evolution of the Kondo scale $T_K$ across the SC phase. The main figure shows $\ln (T_K/\Gamma)$ versus $\tilde{U}'/\tilde{U}$ for  fixed $\tilde U=7$ (upper set of points) and $\tilde{U} =10$ (lower). For
$\tilde{U} = 7$, the inset shows the Kosterlitz-Thouless behaviour close to $\tilde U'_c\simeq 7.046$: here, $\ln (T_K/\Gamma)$ versus $(\tilde U'_c-\tilde U')^{-1/2}$ approaches the asymptotic form of (\ref{eq:ktscale}) with $c\simeq 12.80$ and $
a\simeq 1.38$, shown as a dotted line.}
\end{figure}

As stated in section \ref{sec:thermosc},  we choose to define $T_K$ throughout the SC phase as $T_K=1/\gamma$.
This is most easily calculated using the analytical result (\ref{eq:gamsc}), with the value of $\omega_6$ determined from the numerical energy levels at large $N$. \Fref{fig:kondoscale}
 shows the typical strong-coupling behaviour of $T_K$ across the SC phase, as a function of $\tilde{U}'/\tilde{U}$ for two fixed values of $\tilde{U}$ ($= 7$ and $10$). We see that the SC phase divides naturally into two distinct regimes, namely $\tilde{U}' < \tilde{U}$
and $\tilde{U}' > \tilde{U}$.
For $\tilde U'\le\tilde U$ the Kondo scale increases from the $SU(2)$ value obtaining at $\tilde U'=0$, to a much larger value at the $SU(4)$ point $\tilde{U}'=\tilde{U}$. For $\tilde U'>\tilde U$--- i.e.\ in the charge-Kondo regime ---the scale then decreases very rapidly but continuously with increasing $\tilde U'$, vanishing at the critical point $\tilde U'=\tilde U'_c$. We now consider each regime in turn.

For $\tilde{U}' =0$, the $\tilde{U}$-dependence of the $SU(2)$ Kondo scale 
is of course well known, and our numerical results recover correctly the asymptotic strong coupling form \cite{hewson,kww1}
\begin{equation}
T_K^{SU(2)}\propto\Gamma\tilde U^{\case{1}{2}}\exp\left(\frac{-\pi^2\tilde U}{8}\right).
\end{equation}
On the $SU(4)$ line $\tilde U'=\tilde U$ by contrast we find 
\begin{equation}
T_K^{SU(4)}\propto\Gamma\tilde U^{\case{3}{4}}\exp\left(\frac{-\pi^2\tilde U}{16}\right)
\end{equation}
for large $\tilde{U}$ (including both the exponential and the prefactor), where the reduction in the exponential argument by a factor of $2 \equiv \cal{N}$ is as expected for $SU(2\cal{N})$ Kondo behaviour \cite{hewson}. The marked enhancement of $T_K$ seen in \fref{fig:kondoscale} as $\tilde{U}' \rightarrow \tilde{U}$ thus reflects 
$\tkfour\propto [\tktwo]^{1/2}$ for $\tilde{U} \gg 1$. In fact we see from the figure that $T_K$ grows most rapidly in the close vicinity of the $SU(4)$ point; and
for $U-U'\gg T_K^{SU(4)}$ is not substantially different from its value at $\tilde U'=0$. In later sections we shall see that this is also
reflected in essential persistence
of the $\tilde{U}'=0$ $SU(2)$ spin-Kondo physics over a wide range of $U'$ in the SC phase.

The second SC regime of interest indicated by \fref{fig:kondoscale} is the charge-Kondo regime $\tilde U<\tilde U'<\tilde U'_c$, where the Kondo scale drops rapidly with increasing $U'$  and ultimately vanishes at the phase boundary. The vanishing of $T_K$ is found to be described very accurately by the Kosterlitz-Thouless (KT) \cite{kosterlitz} form
\begin{equation}
\label{eq:ktscale}
T_K=c(\tilde U)\Gamma\exp\left(\frac{-a(\tilde U)}{\sqrt{\tilde U'_c-\tilde U'}}\right)
\end{equation}
where $a$ and $c$ are positive constants for a given $\tilde U$. We note too that (\ref{eq:ktscale}) holds for \emph{all} $U\ge 0$ (and not just strong-coupling),  whence the KT behaviour near the phase boundary is generic. 
The inset to \fref{fig:kondoscale} shows a typical fit to the KT form for $\tilde U=7$, where the constants $a$ and $c$, and the critical $\tilde{U}'_c$, have been chosen to achieve the best possible fit. It is evident from the figure that the numerical results agree very well with equation (\ref{eq:ktscale}) over many orders of magnitude, and that the KT behaviour is in fact observed over much of the charge-Kondo regime. The specific value of $\tilde{U}'_c \simeq 7.046$ thus found also agrees very well with that obtained from the bisection method discussed at the beginning of this section, indicating as such the reliablity of the NRG results.

  In the CO phase by contrast there is no low-energy Kondo scale at all: 
as discussed in the following sections, the low-energy physics here is controlled simply by $U'-U$, which remains finite when the critical $U'_c$ is approached from above.

\subsection{Thermodynamics}
We now turn to a more detailed analysis of the DQD thermodynamics. 
This enables us both to verify the schematic NRG flow diagram shown in \fref{fig:flow}, and to determine the temperature scales at which the flows pass from one fixed point to the next. In the SC phase, it is of course $T_K$ that sets the scale for the ultimate crossover to the SC fixed point. In strong coupling, the low-temperature physics in this phase can thus be described in terms of universal functions of $T/T_K$; we therefore end the section with analysis of this scaling behaviour.

\subsubsection{Entropy.}
\label{sec:entropy}
\begin{figure}
\centering\includegraphics{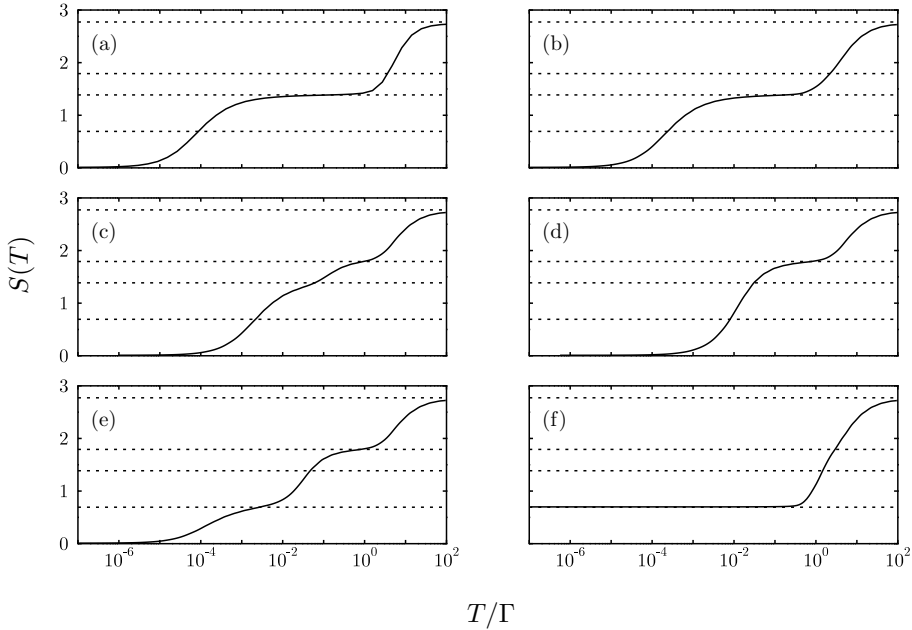}
\caption{\label{fig:entropy} Impurity entropy $S(T)$ versus $T$ for $\tilde U=7$ with (a) $\tilde U'=0$, (b) $\tilde U'=6$ (SC phase, spin-Kondo regime), (c) $\tilde U'=6.9$, (d) $\tilde U'=7$ ($SU(4)$), (e) $\tilde U'=7.03$ (SC, charge-Kondo regime) and (f) $\tilde U'=8$ (CO phase). The critical $\tilde U'_c\simeq 7.046$.
Dotted lines show $\ln$16, $\ln$6, $\ln$4 and $\ln$2, associated with the FO, \lmfour, \lmtwo and CO fixed points respectively.}
\end{figure}

\Fref{fig:entropy} shows $S\equiv S_\mathrm{imp}$ (see \sref{sec:calctherm}) versus $T/\Gamma$ for a range of different interdot interactions $\tilde U'$ at fixed $\tilde U=7$
; we consider points in both the SC ($\tilde U'<\tilde U'_c\simeq 7.046$) and CO ($\tilde U'>\tilde U'_c$) phases. In each case the entropy decreases stepwise with temperature, beginning at the high-$T$ asymptote of $\ln 16$ corresponding to the FO fixed point \cite{kww1}. The plateaus seen as the temperature is lowered (i.e. iteration number $N$ increased, $T \sim D\Lambda^{-N/2}$ \cite{wilson,kww1}) arise when the NRG flow is close to the fixed points of the system, and by identifying the characteristic entropy associated with each fixed point it is possible to deduce the route taken in each case by the NRG flow. 
The reader may find it helpful to refer to the schematic flow diagram of \fref{fig:flow} in combination with the following discussion.

We first recap briefly how the entropy varies with temperature in the well-understood $\tilde U'=0$ limit \cite{kww1}, shown in \fref{fig:entropy}(a). With decreasing temperature it changes from $\ln 16$ to $\ln 4$ ($=2\ln 2$) on the scale $T\sim U$, reflecting the crossover from the FO fixed point to the quadruply-degenerate \lmtwo{} where charge fluctuations on the dots are frozen-out. The entropy then decreases logarithmically slowly with $T$, until at $T\sim \tktwo$, there is a crossover to the SC fixed point where $S_\mathrm{imp} =0$.

How is this changed at finite $\tilde U'$?  
Consider first the $SU(4)$ point $U'=U$, \fref{fig:entropy}(d). Here there is again just a two-stage crossover in the entropy on reducing $T$, but this time the plateau at intermediate-$T$ is at $\ln 6$ rather than the $\ln 4$ seen for $U'=0$; on these temperature scales the physics is controlled by the \lmfour{} fixed point (see \fref{fig:flow}), where all four states of the $(n_L,n_R)=(1,1)$ dot configuration, and the pair of states (2,0) and (0,2), are degenerate. The subsequent crossover from \lmfour to SC occurs on a much higher temperature scale than the analogous crossover from \lmtwo{} to SC in \fref{fig:entropy}(a), consistent with the marked increase in the $SU(4)$ Kondo scale $\tkfour$ over that of $U'=0$.

If one now moves slightly away from $U'=U$, the situation becomes more complicated. On temperature scales $T\gg |U'-U|$, the form of the entropy is found to be the same as at the $SU(4)$ point itself, but now on lower temperature scales there are additional crossovers to the other unstable fixed points of the system. For $U'\lesssim U$, e.g. $U'=6.9$ as shown in \fref{fig:entropy}(c), the NRG flow crosses from \lmfour{} to \lmtwo{} on the scale $T\sim U-U'$, and hence the entropy drops from $\ln 6$ to $\ln 4$. (At lower temperatures still, the local moments of \lmtwo{} are Kondo quenched and the entropy crosses over to the SC value of zero on the scale $T\sim T_K$ as before.)
 On the other hand, if $U'\gtrsim U$, it is the CO fixed point that is approached after \lmfour{}, on the scale $T\sim U'-U$. Here, the remaining dot degree-of-freedom is the charge psuedospin made up of the $(2,0)$ and $(0,2)$ configurations, and the entropy thus plateaus at $\ln 2$. What then happens at lower temperatures depends on the size of $U'$. For $U'<U'_c$, as seen in \fref{fig:entropy}(e) for $\tilde U'=7.03$ ($\tilde U'_c\simeq 7.046$), the CO fixed point is unstable and the flow eventually goes to SC (reflecting the quenching of the charge psuedospin on the temperature scale $T\sim T_K$ (\ref{eq:ktscale})). For $U'>U'_c$ by contrast, 
as shown in \fref{fig:entropy}(f), the CO fixed point is stable and its $\ln 2$ entropy thus persists down to $T=0$. 
\subsubsection{Susceptibilities}
We consider now the impurity spin- and staggered charge-susceptibilities defined at the end of \sref{sec:calctherm}. These probe different aspects of the dot degrees-of-freedom, and show clearly that the transition is from the charge-Kondo state below $U'_c$ to the doubly-degenerate charge ordered state above $U'_c$, the dot spin degrees-of-freedom playing essentially no role. 

\begin{figure}
\centering\includegraphics{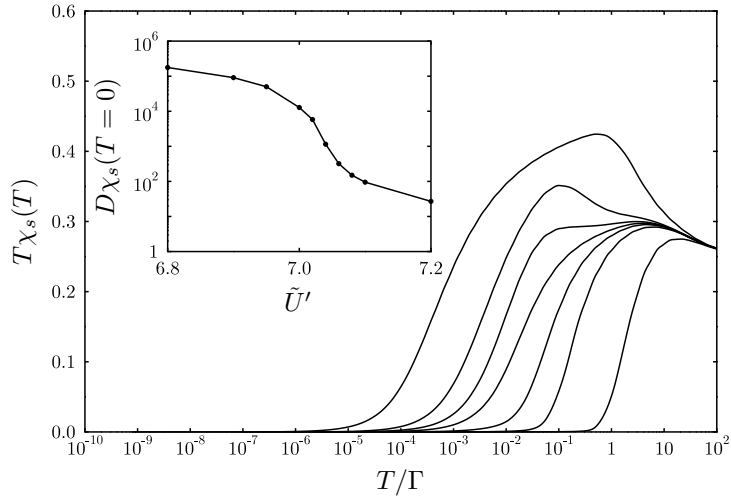}
\caption{\label{fig:chimag} Spin susceptibility $T\chi_s(T)$ versus $T/\Gamma$ for $\tilde U=7$ and $\tilde U'=6$, $6.9$, $6.97$, $7$, $7.03$, $7.1$ and $8$ (from top to bottom). Inset: the zero-temperature $D\chi_s(0)$ versus $\tilde U'$, again for $\tilde U=7$.}
\end{figure}

Figure~\ref{fig:chimag} shows results for the spin susceptibility. 
The main figure shows $T\chi_s$ versus $T$ for a fixed $\tilde U=7$ and a range of $\tilde U'$ spanning the transition ($\tilde{U}'_c \simeq 7.046$): $\tilde U'=6$, $6.9$, $6.97$, $7$, $7.03$, $7.1$ and $8$ (from top to bottom). In all cases the behaviour at high temperatures $(T\gg U)$ is clear: each dot contributes `half a Curie-law' to the spin-susceptibility,  hence $T\chi_s\sim \case{1}{4}$.
For $\tilde U'=6$, on lowering the temperature the form of the susceptibility is seen to be essentially a doubled version of that found for the the $U'=0$ limit \cite{kww1}, the $U'=0$ physics thus persisting far into the SC phase as seen in the previous section
(\emph{cf} e.g. \fref{fig:entropy}(a) and (b)). The susceptibility tends to the full Curie-law value of $\case{1}{2}$ as the NRG flow approaches \lmtwo{}, and then drops to zero at $T\sim T_K$ where the free spins are quenched by the Kondo effect.  

As $U'$ is increased toward $U$, $\chi_s$ begins to show signs of the \lmfour{} fixed point; this is readily shown to have a characteristic $T\chi_s=\case{1}{3}$, and is seen as a shoulder in the $\tilde U'=6.9$ curve as it heads toward to the \lmtwo{} limit of $\case{1}{2}$. At $U'=U$ itself, this shoulder disappears and $T\chi_s$ evolves smoothly to zero with decreasing $T$.

Perhaps the most striking feature of \fref{fig:chimag} is that it shows no sign of the quantum phase transition: the spin susceptibilities evolve smoothly as $\tilde U'$ passes through $\tilde U'_c$ and as $T\to 0$, $T\chi_s$ vanishes in both phases. The reason is as follows. For values of $U'$ much more than $\sim\tkfour$ below $U$, the low-temperature physics consists of the quenching of the \lmtwo{} local moments on the temperature scale $T_K\approx \tktwo$. On increasing $U'$ towards the
$SU(4)$ point $U' =U$, all that happens is that $T_K$ rises slowly (see \fref{fig:kondoscale}) and hence $T\chi_s$ crosses over to zero at progressively higher temperatures -- as seen in \fref{fig:chimag}. Once $U'$ reaches $U$, it is the quenching of the `magnetic' $(n_L,n_R) =(1,1)$ components of the \lmfour{} local-moment states that show up in the \emph{spin}-susceptibility, as a crossover from $\case{1}{3}$ to $0$ at $T \sim \tkfour$. If $U'$ is now increased further, into the
charge-Kondo regime, the magnetic degrees-of-freedom are frozen out at an even higher scale $T \sim U'-U$ (the difference in energy between the $(1,1)$ and $(2,0)/(0,2)$
dot states), corresponding to the crossover from \lmfour to CO. Whether the flow eventually heads to SC at very low temperatures or remains at CO is irrelevant: the magnetic behaviour is already lost at $T\sim U'-U$, and hence the susceptibility evolves smoothly through $U'=U'_c$.

This is further confirmed by the inset to figure~\ref{fig:chimag}, showing the zero-temperature $\chi_s(T=0)$ versus $\tilde{U}'$. We find that 
$\chi_s(0)$ crosses over from a relatively large value indicative of the $SU(2)\times SU(2)$ spin-Kondo physics to a somewhat smaller value at the $SU(4)$ point, and then decreases further still as the charge-Kondo regime is entered and the non-magnetic $(2,0)$ and $(0,2)$ impurity configurations dominate the $T\to 0$ behaviour. Upon moving into the CO phase, the spin-susceptibility simply evolves continuously, and is found to decay smoothly to zero as $U'\to \infty$.

\begin{figure}
\centering\includegraphics{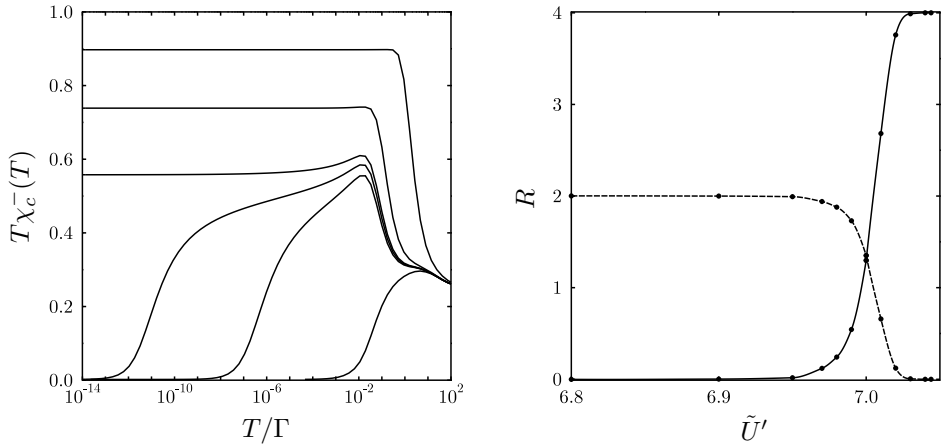}
\caption{\label{fig:chicmin} Left: staggered charge susceptibility. $T\chi_c^-(T)$ versus $T/\Gamma$, for fixed $\tilde U=7$, with $\tilde U'=7$, $7.04$, $7.044$, $7.048$, $7.1$ and $8$ (from bottom to top) and $\tilde{U}'_c \simeq 7.046$. Right: $\tilde U'$-dependence of the Wilson ratios $R_c^-$ (solid line) and $R_s$ (dashed) for $\tilde U=7$.
}
\end{figure}

The situation is quite different for the staggered charge susceptibility $\chi_c^-$
((\ref{eq:chicpmdef})), the charge pseudospin analogue of the spin susceptibility.
\Fref{fig:chicmin} (left) shows 
$T\chi_c^-(T)$ versus $T/\Gamma$ for fixed $\tilde U=7$, with $\tilde U'=7$, $7.04$, $7.044$, $7.048$, $7.1$ and $8$.
Here, the behaviour characteristic of the two phases is qualitatively distinct, with $T\chi_c^-$ vanishing as $T\to 0$ in the SC phase but tending to a finite constant in the CO phase. This reflects the fact that the charge pseudospin comprised of the (2,0) and (0,2) states is screened by the charge-Kondo effect at low-$T$ in the SC phase, but remains free in the CO phase and hence gives rise to a staggered charge susceptibility $\chi_c^-\propto 1/T$. So whereas the spin-susceptibility $\chi_s(0)$ shown in the inset of \fref{fig:chimag} evolves smoothly through the QPT, the
$T=0$ staggered charge-susceptibility diverges as $U'\to U'^{-}_c$, and remains infinite throughout the CO phase. This is directly analogous to the magnetic susceptibility of the two-dimensional $xy$ model \cite{kosterlitz}, the prototype of Kosterlitz-Thouless physics. Note moreover that the divergence of $\chi_{c}^{-}(0)$ tracks precisely the vanishing of the Kondo scale as the transition is approached: from (\ref{eq:wilsondef}) with $T_K \equiv 1/\gamma$, and using $R_c^- \rightarrow 4$ as $U' \rightarrow U'^{-}_{c}$ ((\ref{eq:doodah})), we have $\chi_{c}^{-}(T=0) \sim
3/(\pi^2 T_K)$.

The full $T$-dependences of the staggered charge susceptibilities shown in \fref{fig:chicmin} can likewise be explained in terms of the fixed-point picture used for $\chi_s$.
Here we simply note two points. First, the crossover in $T\chi_c^-$ to zero occurs on the scale $T \sim T_K$ only when $U'\ge U$; for $U'<U$ the charge degrees of freedom are frozen-out on the scale $T\sim U-U'$ instead (cf.\ the discussion of the spin-susceptibility above). Second, the limiting values of $T\chi_c^-$ as $T\to 0$ in the CO phase are clearly $U'$-dependent; this arises naturally from the potential scattering generated at the CO fixed point (and considered further in the Appendix). In the absence of such potential scattering --- as arises in the limit $\tilde{U}'\to\infty$ --- the $T\to 0$ limit of $T\chi_c^-$ is unity.

\subsubsection{Wilson Ratios}
The three Wilson ratios characteristic of the SC phase ($R_s$ and $R_c^\pm$) have
been considered in \sref{sec:thermosc}, and exact results for them obtained at the points $U'=0$, $U$ and $U'^{-}_c$. Here we
show the corresponding numerical results as a function of $\tilde{U}'$; this 
shows the $\tilde{U}'$ scales over which the ratios change from one value to the next, and
also provides an indication of the accuracy of the NRG calculations. \Fref{fig:chicmin} (right) shows the $\tilde U'$-dependence of $R_s$ and $R_c^-$ for $\tilde U=7$; as $\tilde U$ here is large, the uniform charge susceptibility and hence $R_c^+$ is found in practice to be essentially zero over the entire $U'$ range. 
We see that the $R_s=2$, $R_c^-=0$ result known for the $U'=0$ limit \cite{wilson,nozieres} persists with increasing $\tilde{U}'$ until \emph{very} close to the $SU(4)$ point; but then the two Wilson ratios undergo a rapid crossover to the values $R_s=0$ and $R_c^-=4$
((\ref{eq:doodah})), which values obtain in practice over a large part of the charge-Kondo regime. At the $SU(4)$ point itself, we recover numerically (to within $\simeq 2\%$) the exact result $R_s=R_c^-=\case{4}{3}$ ((\ref{eq:doodahday})) expected from the symmetry of the model. 

\subsubsection{Universal scaling behaviour.} Finally, we consider the important issue of the scaling properties of thermodynamics in the SC phase (naturally in strong coupling, $\tilde{U} \gg 1$, where the Kondo scale $T_K$ is exponentially small and well separated from all non-universal scales).
We point out immediately that it is not possible to define a single universal scaling function (for each thermodynamic property) that applies throughout the SC phase. The reason is simple: although the spin-Kondo regime ($U' < U$), $SU(4)$ point ($U' =U$) and charge-Kondo regime ($U < U' < U^{'}_{c}$) all belong to the SC phase, the crossover from one to the next being continuous with increasing $U'$, each has its own characteristic low-temperature behaviour because the particular degrees-of-freedom quenched by the Kondo effect are different in each case. This is particularly apparent from the Wilson ratios shown in \fref{fig:chicmin}. We thus focus on the three distinct regimes separately, beginning with spin-Kondo.

\begin{figure}
\centering\includegraphics{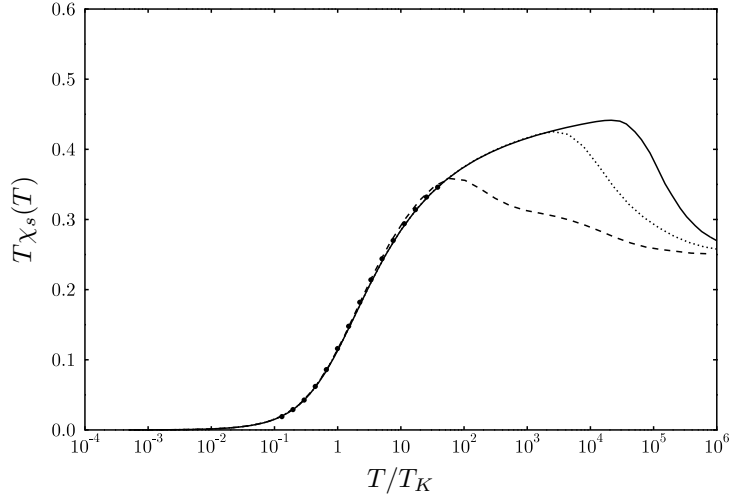}
\caption{\label{fig:chisscal} Universal scaling of the spin-susceptibility in the spin-Kondo regime: $T\chi_s$ versus $T/T_K$ for $\tilde U=7$ and $\tilde U'=0$ (solid line), $6$ (dotted line) and $6.9$ (dashed line). The points plotted in the figure are double the universal single-impurity Kondo susceptibilities, taken from Table V of \cite{kww1}.}
\end{figure}

In the spin-Kondo regime, it is the crossover of the spin-susceptibility $T\chi_s$ on the scale of $T_K$ that characterises the physics of the spin-Kondo effect. In \fref{fig:chisscal}, we show $T\chi_s$ versus $T/T_K$ for $\tilde U'=0$ (solid line), $6$ (dotted line) and $6.9$ (dashed line), all with $\tilde U=7$. It is clear from the figure that the rescaled susceptibilities all collapse onto a common curve at low temperatures (and that the crossover from \lmtwo{} to SC indeed occurs at $T\sim T_K$). Also plotted in \fref{fig:chisscal} (as a set of points) are double the universal \emph{single}-impurity Kondo susceptibilities taken from Table V of \cite{kww1}, which are clearly in excellent agreement with the scaling curves obtained here. [The Kondo scale of \cite{kww1}, here denoted by $T_W$, is easily shown to be related to our definition $T_K=1/\gamma$ by $T_K=3 T_W/(\pi^2 w)$, with $w=0.41071\ldots$ the Wilson number \cite{hewson}]. 

\begin{figure}
\centering\includegraphics{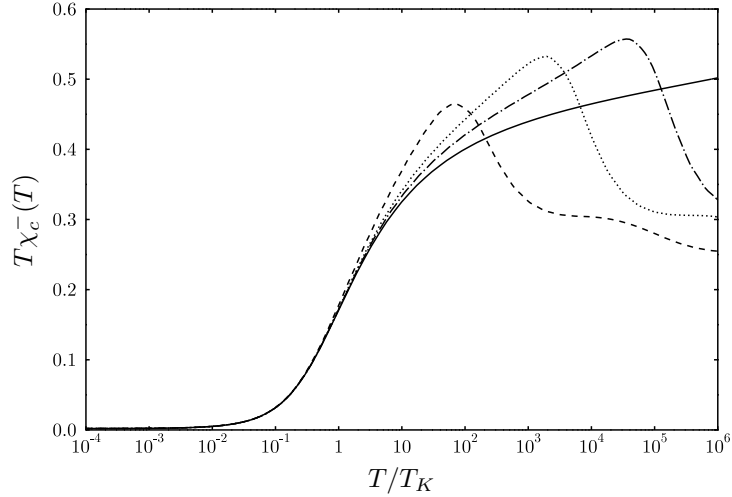}
\caption{\label{fig:chicscal} Universal scaling of the staggered charge-susceptibility in the charge-Kondo regime: $T\chi_c^-$ versus $T/T_K$ for $\tilde U=7$ with $\tilde U'=7.044$ (solid line), $7.04$ (dotted) and $7.03$ (dashed), and for 
$\tilde U=8$ with $\tilde U'=8.028$ (dot-dashed line).}
\end{figure}

In the charge-Kondo regime, it is by contrast the staggered charge-susceptibility that shows universality. In \fref{fig:chicscal} we plot $T\chi_c^-$ versus $T/T_K$ for $\tilde U=7$ with $\tilde U'=7.044$ (solid line), $7.04$ (dotted) and $7.03$ (dashed), and also for $\tilde U=8$ with $\tilde U'=8.028$ (dot-dashed line). That the curves collapse onto a common form at low-temperatures is again evident, the scaling becoming better as $\tilde U'\to\tilde U'^{-}_c$ where the fixed point Hamiltonian (\ref{eq:hnone}) becomes increasingly valid. Points in the spin-Kondo regime ($U'< U$) do not scale onto this universal curve, because there the charge susceptibility vanishes on a finite energy scale $U-U'$ (such that
$(U-U')/T_K\to\infty$ is thus `projected out' in the formal scaling limit $T_K \rightarrow 0$). 

\begin{figure}
\centering\includegraphics{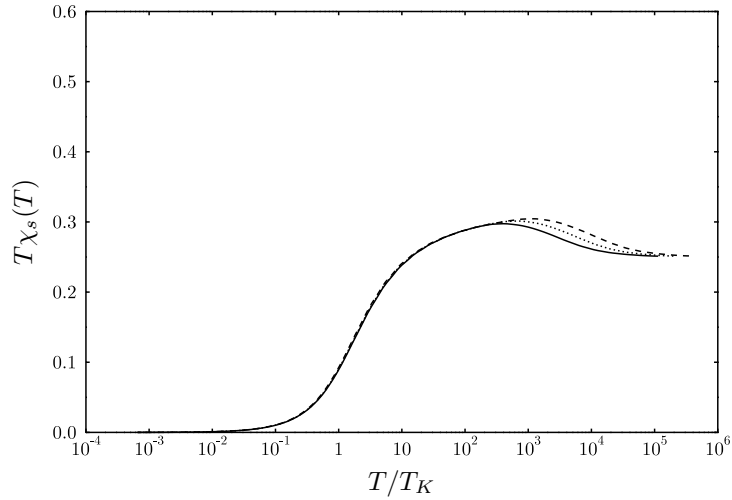}
\caption{\label{fig:chisu4scal} Universal scaling of the spin-susceptibility along the $SU(4)$ line: $T\chi_s$ versus $T/T_K$ for $\tilde U=\tilde U'=7$ (solid line), $8$ (dotted line) and $9$ (dashed line). In this case the spin and staggered charge susceptibilities coincide, $\chi_s = \chi_{c}^{-}$.
}
\end{figure}

Finally, the behaviour along the $SU(4)$ line $U' =U$ is seen in \fref{fig:chisu4scal}
where $T\chi_s$ versus $T/T_K$ is shown for $\tilde U=\tilde U'=7$ (solid line), $8$ (dotted) and $9$ (dashed); and from which universality is again evident.
By exploiting the $SU(4)$ symmetry the behaviour of $T\chi_c^-$ versus $T/T_K$ is \emph{identical} here, and that is indeed recovered by the numerical results. 

\section{Concluding remarks}

Using the numerical renormalization group 
we have considered a symmetrical, capacitively coupled DQD in the two-electron regime, and investigated the evolution of the system as a function of both the interdot and intradot coupling strengths. The range of physical behaviour found is 
both broad and subtle, encompassing a strong coupling, Fermi liquid phase -- itself 
divided into spin-Kondo and charge-Kondo regimes separated by a spin-charge entangled 
$SU(4)$ line -- as well as a broken symmetry charge ordered phase, and the Kosterlitz-Thouless quantum phase transition occurring between the two. 
As far as specific physical properties are concerned, we have focussed largely in this paper on static properties, such as `impurity' thermodynamics, spin- and charge-susceptibilities etc. Importantly, the NRG approach also enables us to determine dynamical properties such as local single-particle spectra~\cite{hewsoncosti}
and related transport properties. It is to these we shall turn in a subsequent paper~\cite{next}. \\

\bf Acknowledgements \rm

We are grateful to the EPSRC for supporting this research, and to the Royal Society for travel support.

\appendix
\section*{Appendix}
\setcounter{section}{1}
In the strongly correlated regime, one can of course obtain a low-energy effective model for the $n=2$ regime by performing a Schrieffer-Wolff transformation \cite{hewson} of (\ref{eq:hnought},{\it b}) to order $V^2$. The final result is
\begin{equation}
\label{eq:heffsw}
\hat H_\mathrm{eff} = \sum_{i,\bk,\sigma}\epsilon_\bk\adkisig\akisig - \Delta\phi(1-\nhat_L\nhat_R) + \hat H_\mathrm{eff}^{(2)},
\end{equation}
where
\begin{eqnarray} 
\label{eq:heff2}
\hat H_\mathrm{eff}^{(2)} &= J\sum_{\substack{i,\bk,\bk',\sigma, \sigma'}}(\cd_{i\sigma'}\cnod_{i\sigma} - \half\delta_{\sigma\sigma'})\ad_{i\bk\sigma}\anod_{i\bk'\sigma'}\nhat_L\nhat_R\nonumber\\
&+ K \sum_{\substack{i,\bk,\bk',\sigma}}(\nhat_i - 1)\ad_{i\bk\sigma}\anod_{i\bk'\sigma}\nonumber\\
&+(\half J + K)\sum_{\substack{i, j\ne i,\bk,\bk',\sigma, \sigma'}}\cd_{j\sigma'}\cnod_{i\sigma}\ad_{i\bk\sigma}\anod_{j\bk'\sigma'}
\end{eqnarray}
and the coupling constants are
\begin{equation}
\label{eq:heff3}
\Delta\phi = U'-U,\qquad J = \frac{4V^2}{U}\qquad\mbox{ and }\qquad K = \frac{2V^2}{2U'-U}.
\end{equation}
The first term on the right-hand side of \eref{eq:heff2} acts solely on states of the $(n_L,n_R) = (1,1)$ dot configuration. It is simply a separable pair of spin-$\half$ Kondo couplings between each dot and its respective lead. By contrast, the second term in \eref{eq:heff2} acts only on states of the $(0,2)$ and $(2,0)$ configurations; this is the correlated potential scattering term that arises at the CO fixed point (see \sref{sec:cofixedpoint}). It is the third term in \eref{eq:heff2} that mixes the different charge sectors, but note that although $(1,1)$ connects to both $(0,2)$ and $(2,0)$ to order $V^2$, the $(0,2)$ and $(2,0)$ states do \emph{not} connect directly to each other. For the special case of $U'=U$, one obtains $K=\half J$ and the Hamiltonian then takes the manifestly $SU(4)$ symmetric form
\begin{equation}
\hat H_\mathrm{eff} = \sum_{\bk,m}\epsilon_\bk\ad_{\bk m}\anod_{\bk m} + J\sum_{\substack{\bk,\bk',m, m'}}(\cd_{m'}\cnod_{m} - \half\delta_{m,m'})\ad_{\bk m}\anod_{\bk' m'}
\end{equation} 
with $m=(i,\sigma)$ as before.

The effective Hamiltonian above leads to a simple physical picture of the DQD physics with increasing $U'$. When $U'=0$, only the first term in \eref{eq:heff2} need be retained: the $(0,2)$/$(2,0)$ states lie an energy $U$ above the $(1,1)$ ground state by virtue of the $\Delta\phi$ term in \eref{eq:heffsw}, and hence do not contribute to leading order in $V^2/U$. The effective Hamiltonian is thus two uncoupled spin-$\half$ Kondo models as one would of course expect. This situation does not change with increasing $U'$ until the energy gained by mixing $(0,2)$/$(2,0)$ into the $(1,1)$ ground state --- of order the $SU(4)$ Kondo temperature $\tkfour$ --- is sufficient to outweigh the cost of promoting electrons into these excited states, $\Delta\phi$. Hence one would expect a rapid crossover from the $SU(2)\times SU(2)$ Kondo effect characteristic of $\tilde U'=0$, to the $SU(4)$ Kondo behaviour of $U'=U$, at a $U'\sim U-\mathcal O(\tkfour)$ as indeed seen in \fref{fig:kondoscale}. Once $U'>U$, the order of the $(0,2)$/$(2,0)$ and $(1,1)$ states is reversed with the latter being the excited state through which the former interconvert (via the third term in \eref{eq:heff2}). If $\Delta\phi$ is sufficiently small, then the interconversion of $(0,2)$ and $(2,0)$ via $(1,1)$ gives rise to a charge-Kondo effect that quenches the
$(2,0)/(0,2)$ charge pseudospin,
and thus the SC phase persists above $U'$ (as shown in \fref{fig:pb} and the associated discussion). Once $U'\gtrsim U+\mathcal O(\tkfour)$ however, the stabilisation energy associated with this charge-Kondo effect cannot compensate for the additional cost of occupying the $(1,1)$ excited states, the $(0,2)$ and $(2,0)$ states do not therefore communicate, and the system undergoes the quantum phase transition to the CO phase.

\begin{figure}
\centering\includegraphics{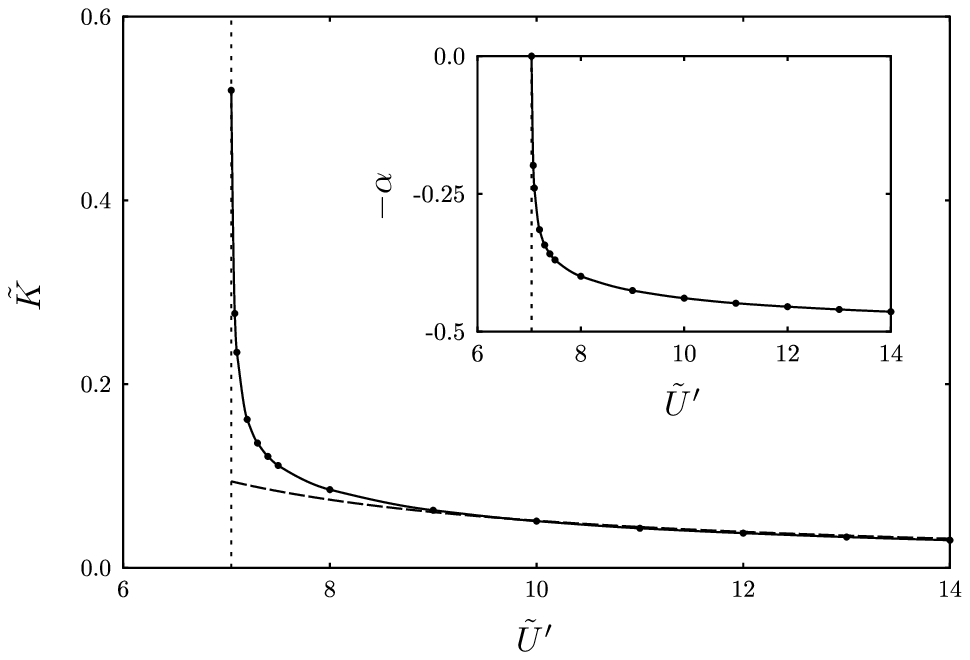}
\caption{\label{fig:kvsut} CO fixed point potential scattering strength, $\tilde K$, vs. $\tilde U'$ for fixed $\tilde U=7$. The dotted line marks the critical $\tilde U'_\mathrm{c}\simeq 7.046$ while the dashed line is the approximate result \eref{eq:kvsut} derived from the $\mathcal{O}(V^2)$ effective Hamiltonian. Inset: the corresponding anomalous exponent $-\alpha$, calculated using \eref{eq:cophaseshift} and \eref{eq:alphadef}.}
\end{figure}

Deep in the CO phase, when $U'\gg U$, one can simplify the effective Hamiltonian further still by neglecting the terms in \eref{eq:heff2} that involve the $(1,1)$ states. This leaves only the correlated potential scattering described by the second term. The resulting effective Hamiltonian in the $(0,2)$/$(2,0)$ manifold is then a continuum form of the CO fixed point Hamiltonian \eref{eq:hncok}, as mentioned in \sref{sec:cofixedpoint}. This means that one can obtain the fixed point Hamiltonian \eref{eq:hncok} by direct discretisation of the effective Hamiltonian; from which the following relationship between $K$ and $\tilde K$ is obtained
\begin{equation}
\tilde K = \frac{4}{1+\Lambda^{-1}}~A_\Lambda\rho K
\end{equation}
with $A_\Lambda=\half(1+\Lambda^{-1})(1-\Lambda^{-1})^{-1}\ln(\Lambda)$ the standard correction factor for the discretisation of the conduction band \cite{kww1}. In terms of the parameters of the full DQD Hamiltonian (\ref{eq:hnought},{\it b}), this becomes 
(using \eref{eq:heff3})
\begin{equation}
\label{eq:kvsut}
\tilde K = \frac{4\ln \Lambda}{\pi^2(1-\Lambda^{-1})}~\frac{1}{2\tilde U' - \tilde U},
\end{equation}
which we plot in \fref{fig:kvsut} (dashed line) as a function of $\tilde U'$ for fixed $\tilde U=7$ in the CO phase. The solid line shows the value of $\tilde K$ obtained from our full numerics, and is indeed seen to fall rapidly onto \eref{eq:kvsut} as one moves deeper into the CO phase. The inset of \fref{fig:kvsut} shows the corresponding anomalous exponent $-\alpha$, calculated from $\tilde K$ using \eref{eq:cophaseshift} and \eref{eq:alphadef}. At the phase boundary itself ($\tilde U'_\mathrm{c}\simeq 7.046$), $\alpha$ vanishes and the perturbation $\delta H_{10}$ in \tref{tab:fixedpointcorrs} is marginal. On moving into the CO phase, $\alpha$ increases, tending to $\half$ in the limit $\tilde U'\to\infty$. Note that $\tilde K$ can of course be obtained by analysing the NRG flows directly; in practice the result is indistinguishable from that obtained using the method described above.

\end{document}